\documentclass[aps,prb,twocolumn,showpacs,floatfix]{revtex4-1}
\pdfoutput=1
\usepackage[T1]{fontenc}
\usepackage[utf8]{inputenc}
\usepackage{graphicx}
\usepackage{amsmath,amssymb,bm,dcolumn,times}
\usepackage{amsmath}
\usepackage{color}
\usepackage{ulem}

\newcommand{\be}{\begin{equation}}
\newcommand{\ee}{\end{equation}}
\newcommand{\bea}{\begin{eqnarray}}
\newcommand{\eea}{\end{eqnarray}}

\begin{document}

\title{Pumping electrons in graphene to the $\mathbf{M}$-point in the Brillouin zone: The emergence of anisotropic plasmons}

\author{A. J. Chaves}
\email{andrej6@gmail.com}
\affiliation{Department of Physics and Center of Physics, University of Minho, P-4710-057, Braga, Portugal}

\author{N. M. R. Peres}
\email{peres@fisica.uminho.pt}
\affiliation{Department of Physics and Center of Physics, University of Minho, P-4710-057, Braga, Portugal}

\author{Tony Low}
\email{tlow@umn.edu}
\affiliation{Department of Electrical and Computer Engineering, University of
Minnesota, Minneapolis, Minnesota 55455, USA}

\date{\today}

\begin{abstract}
We consider the existence of plasmons in a non-equilibrium situation where electrons from the valence band of graphene
are pumped to states in the Brillouin zone around the $\mathbf{M}$-point  by a high intensity UV electromagnetic field.
The resulting out-of-equilibrium electron gas is later probed by a weak electromagnetic field of different frequency. 
We show that the optical properties of the system and the dispersion of the plasmons are strongly anisotropic,
depending on the pumping radiation properties: its intensity, polarization, and frequency.
This anisotropy has its roots in the saddle-like nature of the electronic dispersion relation
around that particular point in the Brillouin zone. 
 It is found that despite the strong anisotropy,
the dispersion of the plasmons scales with the square root of the wave number but is characterized an effective Fermi energy, which 
depends on the properties of the pumping radiation. Our calculations go beyond the usual Dirac cone approximation taking the full band structure of graphene
into account. This is a necessary condition for discussing plasmons at the $\mathbf{M}$-point in the Brillouin zone.
\end{abstract}

\maketitle

\section{Introduction}

\begin{figure}  
   \vspace{0.5cm}
   \centering
   \includegraphics[scale=0.5]{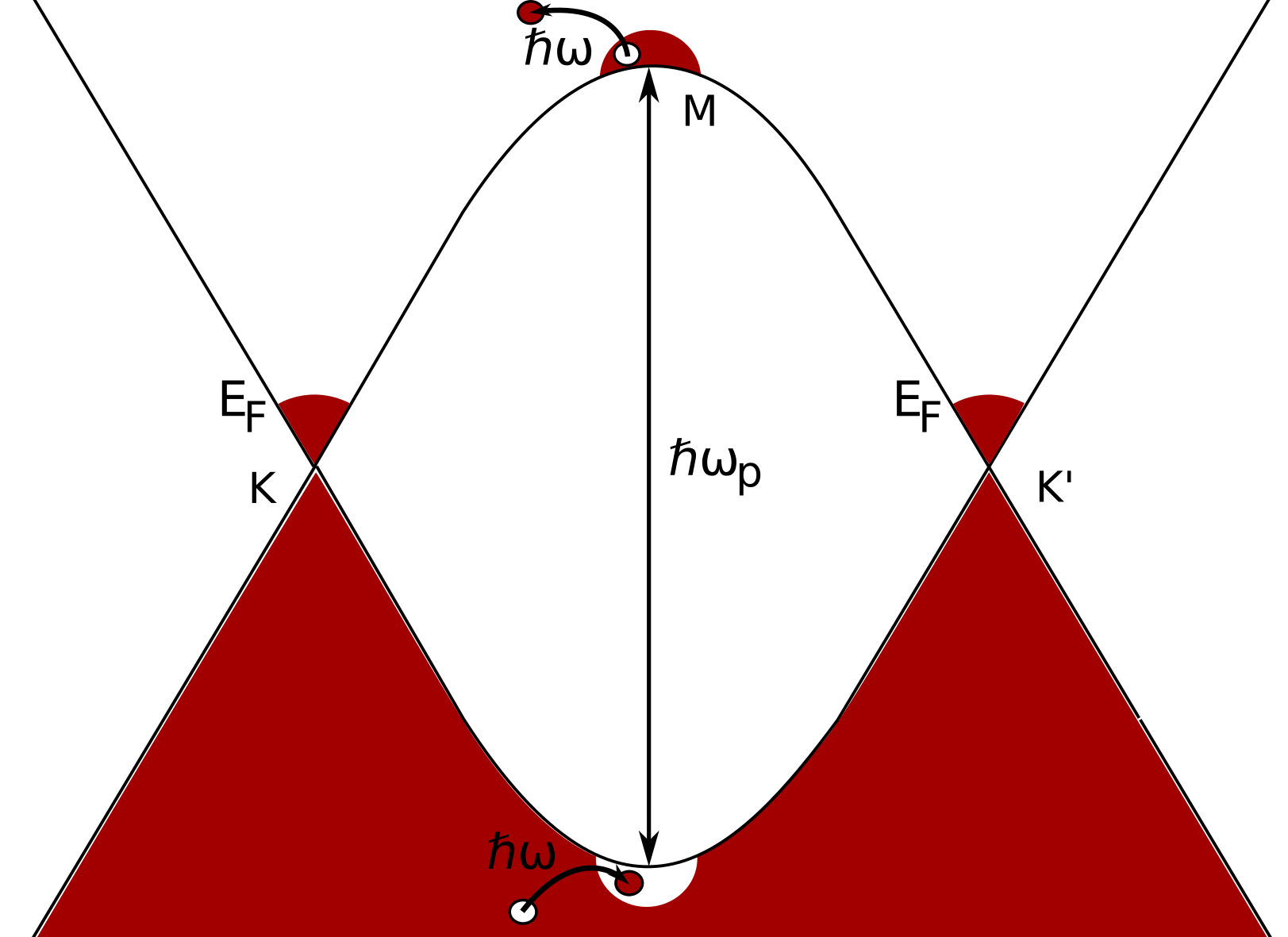}
   \caption{(Color on-line.) 
   This figure represents the physical situation we are considering in this paper: a possibly doped graphene is driven out-of-equilibrium by an electric field of frequency $\omega_p$ that creates an electron gas around the
   $\mathbf{M}$-point in the Brillouin zone. 
The created plasma is then probed by a field of frequency $\omega\ll\omega_p$.   
   We note that the mechanism we are discussing does not require an initially doped graphene, that is, we can have $E_F=0$. Also note that the electronic spectrum at the 
   $\mathbf{M}$-point has a saddle-like nature (in the drawing only the 
   steepest descent direction is shown; see Fig. \ref{fig_full_spectrum} for a drawing of the full band structure). The probe field allows the excitation of an electron,
   belonging to the out-of-equilibrium gas,
   to higher energies.}
   \label{intr}
\end{figure}

Light matter interaction at low energies occurs due to the interaction of electromagnetic radiation with 
weakly bound
electrons in a material. The
electron gas in a conductor  is a well known form of weakly bound electrons which couple to electromagnetic radiation. Usually, we assume that the electron gas is
in equilibrium and that the external perturbation is small, in which case the response of the gas can be computed in terms of its equilibrium properties ---linear response theory. The situation is markedly different when a high-intensity electromagnetic radiation interacts with an electron gas driving it to an out-of-equilibrium regime. In this case
the response of the system depends on intensity of incoming radiation and on the orientation of its polarization relatively to the 
real space lattice of the crystal. Moreover the distribution function of the electron gas occupancy is no longer a Fermi-Dirac distribution. It is this situation that will be studied in this paper.  Here we consider an intense pumping electromagnetic field interacting with the weakly bound electrons in graphene thus generating an
out-of-equilibrium gas. The pumping is followed by
a weak-probe electromagnetic-field, of frequency much smaller than that of the pumping field, which probes the out-of-equilibrium plasma created by the pumping. The physics
of this process is depicted in Fig. \ref{intr}.

Pumping graphene with electromagnetic radiation is a possible tool to study the dynamics of the charge carriers in graphene. In the work of George {\it et al.} \cite{George2008}  the recombination dynamics and carrier relaxation in graphene was studied with terahertz spectroscopy. 
When electrons in graphene are  pumped by an intense light field pulse, the response is highly anisotropic as was shown in Ref. [\onlinecite{Mittendorff2014}]. After about 1ps of the initial pumping pulse, the photogenerated electrons are described by an isotropic Fermi-Dirac distribution with a high temperature\cite{Gierz2013}. The graphene optical properties under such conditions were studied by Malic {\it et al.} \cite{Malic2011} and Sun {\textit et al.} \cite{Sun2012}. 
The electron dynamics of photo-excited electrons, including the stimulated electron-hole recombination, was studied by Li {\it et al.}\cite{Li2012}.
The plasmon dispersion
relation under a non-equilibrium hot Fermi-Dirac distribution was studied very recently by Page {\it et al.} \cite{Page2015}     and experimentally measured by Ni {\it el al.} \cite{Ni2016}. We note that plasmons in  graphene also offer a possible decay channel to cool down the hot electron gas \cite{Hamm2016}. The optical conductivity of doped and gapped graphene taking into account a non-equilibrium distribution and interband processes was studied theoretically by Singh {\it et al.}\cite{Agarwal2016}. All these studies 
where made in a regime where the validity of the Dirac cone approximation holds, that is, the pumping
field has a frequency in the IR/Vis region of the electromagnetic spectrum.

For a pulsed  laser beam with a pulse duration much larger than 
1 ps ---the case we will consider in this work---, the carrier distribution will remain anisotropic for the duration of the pulse. In this case,   electron-phonon and electron-electron interactions, as
well as the effect of disorder can be encoded into a hot carrier relaxation rate. Under such conditions the surface plasmon-polariton (SPP) was studied in systems described by a gapped Dirac equation by Kumar {\it et al.}\cite{Anshuman2016}, who discussed the response of the electron gas to a circular polarized light. In this approach,  the density matrix equations of motion are solved to  determine the non-equilibrium electronic distribution. In this paper we study the electronic distribution and the plasmon spectrum of an optically pumped graphene. The material is subjected to  a linearly polarized light beam, with a frequency that creates an electron gas beyond the regime where the Dirac approximation holds.
This  is relevant when graphene is subjected to UV radiation. In this case, the spectrum is no longer Dirac-like and the full band structure of the system has to be taken into account. Therefore, this work
goes beyond that of Anshuman {\it et al.}\cite{Anshuman2016}  and includes also the
regime studied by these authors.

The plasmons in graphene were first probed in real-space in the studies of  Fei {\it et al.} \cite{Fei2012} and Chen {\it et al.} \cite{Chen2012}.   In the work [\onlinecite{Ni2016}] the plasmons in graphene are generated by an infrared beam focused in the metalized tip of an atomic force microscope, after a first pumping pulse of electromagnetic radiation. The tip is also used to detect the plasmons that propagate along the graphene surface and after reflection in the sample edges standing waves are 
produced. This kind of experiment can be used to detect the plasmons discussed in the present work using  a pulsed laser
in the UV range (pulse duration much larger than 1 ps) for pumping
the electrons in graphene. Due to excitonic effects\cite{Geim_excitonic} the position of the maximum of absorption associated  to inter-band
transitions at  the
$\mathbf{M}-$point  is reduced from $5.4$ eV (independent-electron result) 
to about $\sim4.6$ eV ($\lambda\sim270$ nm), a wavelength
for which there are available lasers (see also Sec. \ref{fc}). 


Under intense and long optical excitation, the carrier distribution maintains  a non-equilibrium state and does not follow the Fermi Dirac distribution.
The new electronic distribution has to be calculated
using the von-Neumann equation of motion, with a phenomenological relaxation-term that tends to drive the system towards thermal equilibrium,
characterized by a 
Fermi-Dirac distribution. 
Since we want to discuss plasmons in the non-equilibrium electron gas created around the $\mathbf{M}$-point in the Brillouin zone, we need to describe the $\pi-$electrons in graphene using a tight-binding Hamiltonian. 
For graphene in the tight-binding approximation, we have a two-band (valence and conduction) system labeled by
the crystal momentum $\mathbf{k}$, which runs over 
the full hexagonal Brillouin zone. In our calculations, carrier scattering is accounted for via a relaxation rate. 
As such, the resulting
equations need to be solved for each point in the first Brillouin zone and different momentum values  are not explicitly coupled. 

The paper is organized as follows: in Sec. \ref{neqdm} we derive the Bloch equations for the 
electrons in graphene within the tight-binding model, which is 
valid beyond the Dirac cone approximation. 
We study the transient response under a pulse laser and obtain analytical expressions for the out-of-equilibrium electronic distribution. In Sec. \ref{itfpe} we obtain the equations to calculate the out-of-equilibrium susceptibility from the new electronic distribution. In Sec. \ref{lwl} we derive a semi-analytical formula, valid in the long-wavelength regime, for the susceptibility and in Sec. \ref{tacgup} the optical conductivity is obtained in the same conditions. In section \ref{nr} we compute numerically the susceptibility of the out-of-equilibrium electron gas and obtain results for the plasmon dispersion and the loss function.
In Sec. \ref{spp} we use the semi-analytical equations for the conductivity tensor to numerically calculate the relation dispersion of the surface plasmon-polariton and discuss its anisotropic properties.
In Sec. \ref{fc} we provide 
a discussion on the feasibility of an experiment to observed the predicted
effects and 
the general conclusions of  our work.

\section{Non-equilibrium Density-Matrix and Bloch equation in the tight-binding approximation} \label{neqdm}

We consider  the electrons in graphene described by a tight-binding Hamiltonian $H_0$, with a nearest neighbors
hopping
 term only, and subjected to an external electric
field $\boldsymbol{\mathcal E}$ (see Appendix \ref{app_tb}):
\be
H =H_0+V= \sum_{n} t_\text{TB}\left(\hat{a}^\dagger_n \hat{b}_n+\hat{b}^\dagger_n \hat{a}_n \right)+e\boldsymbol{\mathcal E} \cdot\mathbf{R}, \label{H}
\ee
where 
the index $n$ extends over all unit cells of the crystal, 
$t_\text{TB}=2.7$ eV is the hopping parameter, $\hat{a}^\dagger_n$($\hat{b}^\dagger_n$) creates an electron in the site $n$ of the sublattice A(B), and $\mathbf{R}=\mathbf{R}_A+\mathbf{R}_B$ is the position operator:
\begin{subequations}
\bea
\mathbf{R}_A&=&\sum_n \mathbf{R}_n \hat{a}^\dagger_n\hat{a}_n,\\
\mathbf{R}_B&=&\sum_n \left(\mathbf{R}_n+\boldsymbol{\delta}_1\right) \hat{b}^\dagger_n\hat{b}_n,
\eea
\end{subequations}
where
$\boldsymbol{\delta}_1=a_0(1,0)$ is the vector connecting the $A$ and $B$ sub-lattices in the same primitive cell, and $\mathbf{R}_n=n_1 \mathbf{a}_1+n_2\mathbf{a}_2$, $n_1$ and $n_2$ are integers numbers, and the primitive vectors are defined as: $\mathbf{a}_1=a_0(3/2,\sqrt{3}/2)$ and $\mathbf{a}_2=a_0(3/2,-\sqrt{3}/2)$, where
$a_0\approx0.14$ nm is the carbon-carbon distance in graphene. 

The eigenvectors of $H_0$, satisfying the eigenvalue equation 
$H_0|\mathbf{k},\lambda\rangle=\lambda E_\mathbf{k}|\mathbf{k},\lambda\rangle$, are:
\be
|\mathbf{k},\lambda\rangle=\frac{1}{\sqrt{2}}\left(|A,\mathbf{k}\rangle+\lambda e^{i\Theta_\mathbf{k}}|B,\mathbf{k}\rangle \right), \label{eigh0}
\ee
with $\lambda=\pm1$ and  $|A,\mathbf{k}\rangle=\sum_n e^{i\mathbf{k}\cdot\mathbf{R}_n} \hat{a}^\dagger_n |0\rangle$ 
(the same holds for the states $|B,\mathbf{k}\rangle$ upon the replacement $\{a,A\}\leftrightarrow \{b,B\}$). We define the useful auxiliary functions $\phi_\mathbf{k}$ and $\Theta_\mathbf{k}$ through the relation:
\be
\phi_\mathbf{k}=\sum_i e^{i \mathbf{k}\cdot\boldsymbol{\delta}_i}=|\phi_\mathbf{k}|e^{i\Theta_\mathbf{k}}, \label{phifunc}
\ee
with the positive eigenvalues given by $E_\mathbf{k}=t_\text{TB}|\phi_\mathbf{k}|$ and where $\boldsymbol{\delta}_i$ are the vectors connecting an atom $A$ to all the three nearest 
neighbor $B$ atoms [see Eq. (\ref{eq_deltas})].
From now on, all the momenta are measured in units of the inverse of the lattice parameter $a_0$, thus we make the replacement	 $\mathbf{k}\rightarrow\mathbf{k}a_0$. 

The pumping of the electrons in graphene changes the electronic density. Thus the system is described by a non-equilibrium ---but stationary--- distribution. To calculate the system properties in an out-of-equilibrium regime we use the density matrix $\rho(t)$ formalism, whose time-evolution obeys the von-Neumann equation\cite{Fano1957}:
\be
i\hbar \,\partial_t \rho(t)=[H,\rho(t)]. \label{heq}
\ee

We assume a time-dependent and uniform electric field (that is, with null in-plane wave number). Thus the interaction term in the Hamiltonian does not
 couple electronic states
from different points in the Brillouin zone. We then project Eq. (\ref{heq}) into the eigenvectors of $H_0$ given by Eq. (\ref{eigh0}) with
the same wave number $\mathbf{k}$ and different bands. The diagonal part of the density matrix corresponds to the new distribution functions:
\bea
n_{c,\mathbf{k}}(t)=\langle \mathbf{k}, +| \rho(t)| \mathbf{k},+\rangle,\\
n_{v,\mathbf{k}}(t)=\langle \mathbf{k}, -| \rho(t)| \mathbf{k},-\rangle,
\eea
and the off-diagonal elements correspond to transitions probabilities:
\bea
p_{cv,\mathbf{k}}(t)=\langle \mathbf{k}, +|\rho(t)|\mathbf{k}, -\rangle,\\
p_{vc,\mathbf{k}}(t)=\langle \mathbf{k}, -|\rho(t)|\mathbf{k}, +\rangle.
\eea

After the calculation of the commutator in Eq. (\ref{heq})   and the introduction of two phenomenological
relaxation-rate terms responsible for relaxing the non-equilibrium electron gas back to its equilibrium Fermi-Dirac distribution, we obtain a set of coupled equations:
\begin{subequations}
\bea
-\partial_t n_{c,\mathbf{k}}=\gamma_0\left(  n_{c,\mathbf{k}}-f_{c,\mathbf{k}}\right)+i\Omega_\mathbf{k}(t)\Delta p_\mathbf{k},
\\
-\partial_t n_{v,\mathbf{k}}=\gamma_0\left(  n_{v,\mathbf{k}}-f_{v,\mathbf{k}}\right)-i\Omega_\mathbf{k}(t)\Delta p_\mathbf{k},
\\
\left(\partial_t+i\omega_\mathbf{k}+\gamma_p\right) p_{cv,\mathbf{k}}=-i\Omega_\mathbf{k}(t)\Delta n_\mathbf{k},
\\
\left(\partial_t-i\omega_\mathbf{k}+\gamma_p\right) p_{vc,\mathbf{k}}=i\Omega_\mathbf{k}(t) \Delta n_\mathbf{k},
\eea\label{alleq}\end{subequations}
where the interband relaxation rate is represented by $ \gamma_p$ and intraband one by $\gamma_0$; also we have
$\hbar\omega_\mathbf{k}=2E_\mathbf{k}$, $\Delta n_\mathbf{k}=n_{c,\mathbf{k}}-n_{v,\mathbf{k}}$, $\Delta p_\mathbf{k}=p_{cv,\mathbf{k}}-p_{vc,\mathbf{k}}$, and $f_{c/v,\mathbf{k}}$ is the equilibrium Fermi-Distribution for the conduction/valence band. The time dependence of $n_{c/v,\mathbf{k}}$, $p_{vc/cv,\mathbf{k}}$, and $\boldsymbol{\mathcal{E}}$ has been omitted for simplicity of notation, and finally  the Rabi frequency $\Omega_\mathbf{k}$ is given by:
\be
\Omega_\mathbf{k}(t)=\frac{e a_0\boldsymbol{\mathcal E}(t)\cdot \boldsymbol{\nabla}_\mathbf{k} \Theta_\mathbf{k}}{2\hbar}, \label{rabi}
\ee
and couples the diagonal to the off-diagonal elements of the density matrix through the external pumping 
electric field $\boldsymbol{\mathcal E}(t)$. 

\begin{figure}[h] 
   \vspace{0.5cm}
   \centering
   \includegraphics[scale=0.43]{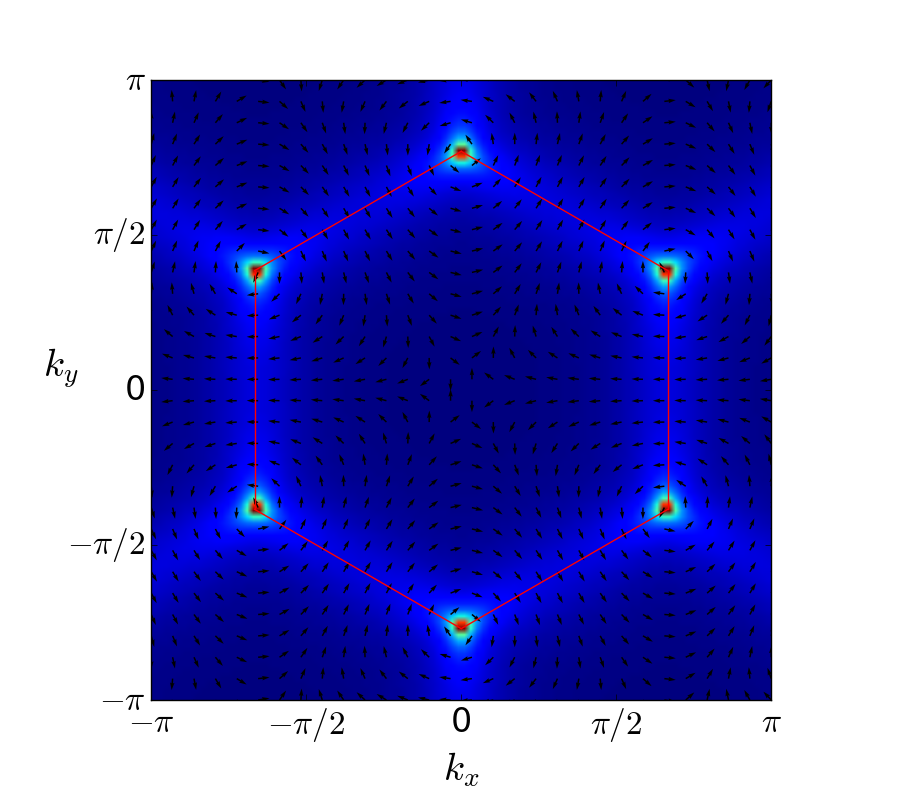}
   \caption{(Color on-line.) Plot of the vector field $\boldsymbol{\nabla}_\mathbf{k} \Theta_\mathbf{k}/||\boldsymbol{\nabla}_\mathbf{k} \Theta_\mathbf{k}||$. This field controls the Rabi frequency and it can be probed by the polarization of the pumping field. Note that the rotation of the vector field
   in the two non-equivalent Dirac points has opposite senses.
   The red hexagon represents  the first Brillouin zone and the intensity
   refers to the absolute value of the vector field $\boldsymbol{\nabla}_\mathbf{k} \Theta_\mathbf{k} $; it is more intense around the Dirac points (brighter spots) and along the directions connecting two Dirac points and passing through the $\mathbf{M}-$point.}
   \label{quiver}
\end{figure}

Equations (\ref{alleq}) are the Bloch equations in graphene\cite{Winzer2013}, with only interband contributions
included (note that we want to excite electrons deep in the valence band to high up in the condution band), and describe the evolution of the electronic distribution and the rate of
interband transitions when an external intense and highly energetic electric field $\boldsymbol{\mathcal E}$ is applied. The vector field $\boldsymbol{\nabla}_\mathbf{k} \Theta_\mathbf{k}$ entering in the Rabi frequency does not
depend on the external parameters and is shown in Fig. \ref{quiver}. The two inequivalents Dirac points are located at the corners of the hexagon in this figure. 
Near these points the function $\Theta_\mathbf{k}$ becomes the angle between the momentum $\mathbf{k}$ and the $x$-axis.

We now comment  on the possible values of $\gamma_0$
(intraband scattering rate; note that this controls what is quantum optics is called the {\it population})
 and $\gamma_p$ (interband scattering rate; note that this controls
 what in quantum optics is called the {\it coherence}).
  Let us first remark that
these quantities have  been scarcely studied in the UV range \cite{Shang2010,Roberts2014,Oum2014}; probably the most
comprehensive study is that of Oum {\it et al.}\cite{Oum2014}. 
Roberts {\it et al.}\cite{Roberts2014} suggest an intraband electron-electron scattering time $\tau_{e-e}<2$ ps (the interband electron-electron scattering time has not been measured), followed by an electron-phonon
intraband scattering time larger than $\tau_0>2$ ps. Theoretically,
these scattering times have not yet been studied at the $\mathbf{M}-$point. 
On the other hand, the carrier dynamics is much better studied when the charge carriers are excited with IR/Vis radiation.  
The dynamics after the initial pulse time of duration $\Delta t_p$
is the following: intraband electron-electron collisions leads the 
out-of-equilibrium electron gas to a quasi-thermal and transient distribution  after a time $\tau_\text{th}$. This distribution is characterized by a temperature $T_\text{el}$ that essentially controls the broadening of the 
optical conductivity features at high frequencies. For longer times
 the system relaxes towards thermal equilibrium via electron-phonon
 coupling, occurring in a time scale $\tau_C$ (cooling time scale), finally  recombination of electron-hole pairs, occurring in a time scale $\tau_R$, takes place. For a pumping-field of photon energy
 $\hbar\omega_p=1.6$ eV George {\it et al.}\,\cite{George2008} identified  three time scales: rapid thermalization, $\tau_{e-e}\sim10-150$ fs (smaller than 60 fs, according to Ref. [\onlinecite{Shang2010}], and around 10 fs,
 according to Ref. [\onlinecite{Giovanni:15}], measured using a $Z-$scan technique), followed by carrier cooling on a time scale of 
$\tau_C\sim0.15-1$ ps ($\hbar/\tau_C\sim4-27$ meV; $\tau_C\sim0.18$ ps, according to Ref. [\onlinecite{Shang2010}]), and finally carrier recombination takes place
characterized by a time scale 
$\tau_R\sim1-15$ ps (for small doping, theoretical estimations give $\tau_R>1$ ps\,\cite{Rana2007}). 
 We associate the time 
 $\tau_C$ with $1/\gamma_0$, with   $\tau_C\sim0.15-1$ ps. The assignment of 
 $\gamma_p$ to a given relaxation rate is more difficult.  Intuition dictates that $\tau_p$ should be determined 
 by interband electron-electron interactions, a quantity not easily
 accessible via pump-probe experiments.
 We therefore concluded that $\tau_R$ in the IR/Vis/UV range is largely
 undetermined at  present. Luminescence studies of graphene irradiated with Vis/UV electromagnetic pulses (30 fs duration) found 
 \cite{Heinz2010}
 a characteristic emission time of $\tau_\text{em}\sim0.01-0.1$
 ps ($\hbar/\tau_\text{em}\sim40-400$ meV). Assuming that 
 the luminescence transition is controlled by $\gamma_p$
 we can consider the longer time of 0.1 ps to estimate
 $\hbar\gamma_p\sim40$ meV (in the figures we shall use 
 $\hbar\gamma_p=2\gamma_0=28$ meV). 

\subsection{Real Time Analysis}

We can simplify the set of Eqs. (\ref{alleq}), defined in terms of two real ($n_{c,\mathbf{k}}$, $n_{v,\mathbf{k}}$) and two complex
$p_{vc,\mathbf{k}}$, $p_{cv,\mathbf{k}}$ quantities, to three equations involving real quantities 
only. We will show that under an intense monochromatic wave, the
electronic density can reach a steady-state.

Using Eqs. (\ref{ax1e}) and  (\ref{ax2e}), we define the deviation from the equilibrium density as $\rho_\mathbf{k}(t)$:
\begin{subequations}
\bea
n_{c,\mathbf{k}}(t)=f_{c,\mathbf{k}}+ \rho_\mathbf{k}(t),
\\
n_{v,\mathbf{k}}(t)=f_{v,\mathbf{k}}- \rho_\mathbf{k}(t),
\eea \label{rhodef}\end{subequations}
and we split the transition rate into real and imaginary parts:
\be
p_{vc,\mathbf{k}}(t)=x_\mathbf{k}(t)+iy_\mathbf{k}(t),
\ee
where $x_\mathbf{k}(t)$ and $y_\mathbf{k}(t)$ are real and $p_{cv,\mathbf{k}}(t)=x_\mathbf{k}(t)-iy_\mathbf{k}(t)$.
From Eqs. (\ref{alleq}) (see Appendix \ref{aprtime}), we  can write:
\begin{subequations}
\bea
 \dot{x}_\mathbf{k}&=&-\gamma_p x_\mathbf{k}  -\omega_\mathbf{k}y_\mathbf{k},
\\
 \dot{y}_\mathbf{k}&=&\omega_\mathbf{k}x_\mathbf{k}-\gamma_p y_\mathbf{k} -\Omega_\mathbf{k}(t)\left(2\rho_\mathbf{k}+ \Delta f_\mathbf{k}\right),
\\
\dot{\rho}_\mathbf{k}&=&-\gamma_0  \rho_\mathbf{k} +2\Omega_\mathbf{k}(t) y_\mathbf{k},
\eea \label{eqtrans}\end{subequations}
where the time dependence in $x$, $y$ and $\rho$ has been omitted and $\Delta f_\mathbf{k}=f_{v,\mathbf{k}}-f_{c,\mathbf{k}}$. Also note the different signs in front of $\rho_\mathbf{k}(t)$
in Eqs. (\ref{rhodef}).

\begin{figure}[h]
   \vspace{0.5cm}
   \centering	
   \includegraphics[scale=0.5]{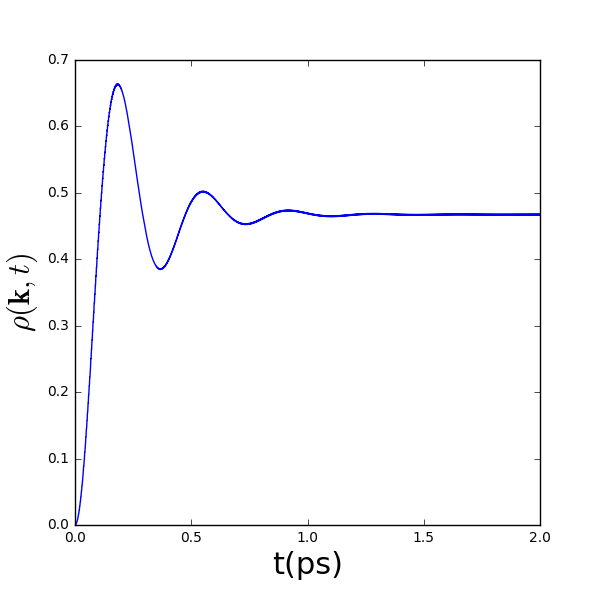}
   \caption{(Color on-line.) Plot of the time evolution of the electronic distribution $\rho_\mathbf{k}(t)$ for $\mathbf{k}=2\pi/a_0(3,0)$, $\hbar\omega_p=2t_\text{TB}$,
   $\tau_0=300$ fs (see Ref. [\onlinecite{Winnerl2011}]),
   which corresponds to
   $\hbar\gamma_0=14$ meV (see also Ref. [\onlinecite{Wang2011}]), $\hbar\gamma_p=28$ meV,  $\mathcal{E}_0=0.5 $ GV/m, and $E_F=0.2$ eV.
   The pumping field is linearly polarized along the $x$-axis. The steady state is attained after 1 ps.}
   \label{rtime} 
\end{figure}

The set of coupled Eqs. (\ref{eqtrans}) describe, using three real functions $x$, $y$, and $\rho$, 
both the time evolution of the transition probability and the electronic density in the reciprocal space. Since we have included the effect  of both electron-electron and  electron-phonon interactions using only a constant relaxation rate, there is no coupling between excitations from different $\mathbf{k}$. Thus, for each point in the
Brillouin zone we can solve Eqs. (\ref{eqtrans}). In Fig. \ref{rtime} we plot the time evolution of the function $\rho_\mathbf{k}$, for $\mathbf{k}=2\pi/a_0(3,0)$ (that corresponds to the $\mathbf{M}$-point in the Brillouin zone) for a monochromatic electric field of frequency $\hbar\omega_p=2t_\text{TB}$ (that corresponds to a vertical transition at the $\mathbf{M}-$points) and intensity $\mathcal{E}_0=0.5 $ GV/m (this is a moderate field intensity), with linear polarization along the x-axis.

\begin{figure*}[ht]
   \centering
   \includegraphics[scale=0.35]{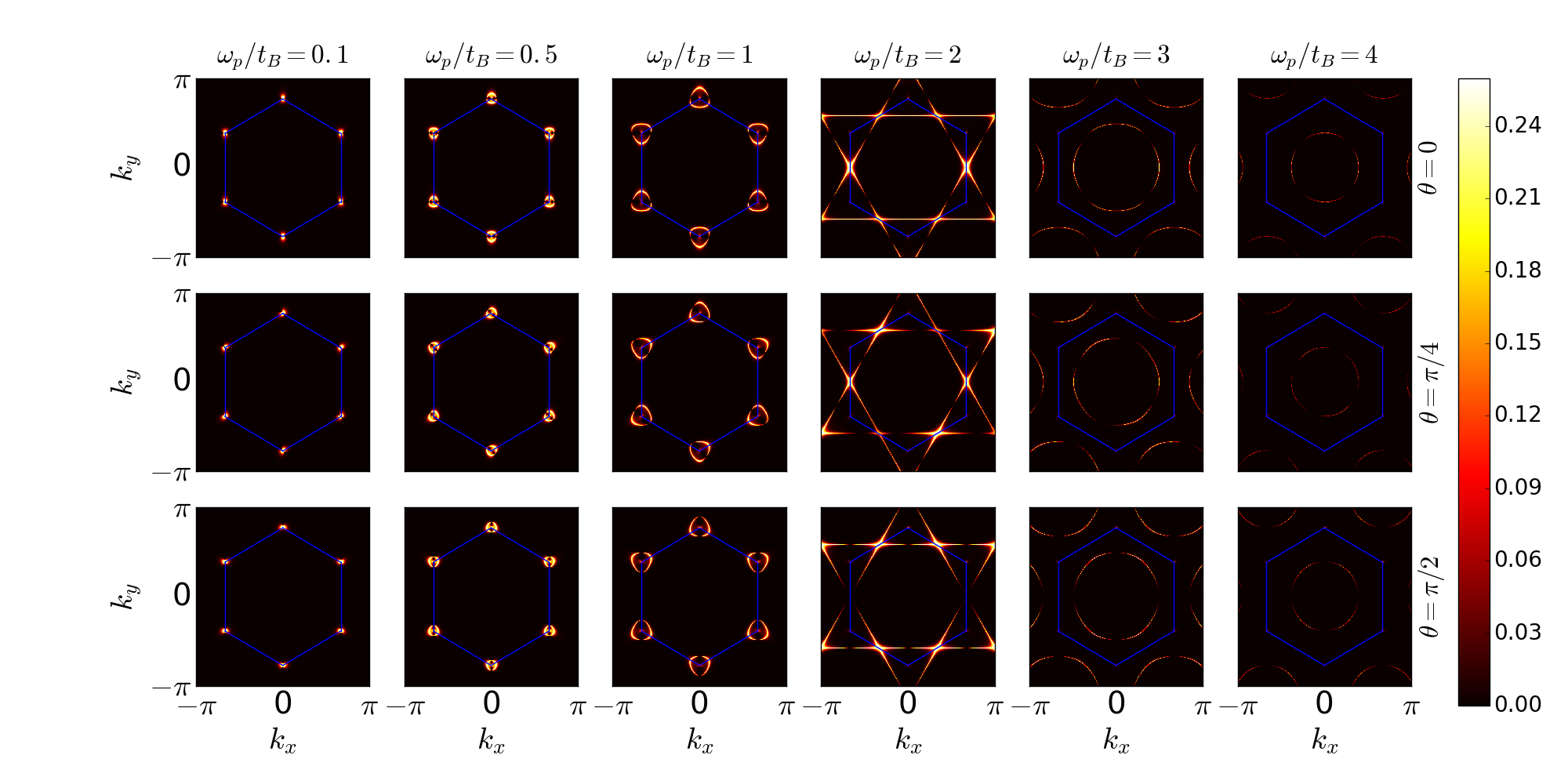}
   \caption{(Color on-line.) Plot of the electronic density $\rho_\mathbf{k}$ for different values of the pumping frequency $\omega_p$ and pumping orientation $\theta$ with intensity $\mathcal{E}_0=0.5 $ GV/m, $E_F=0$, 
   $\hbar\gamma_0=14$ meV, and $\hbar\gamma_p=28$ meV. The bright regions in the Brillouin zone depend on the  value of $\omega_p$ and have the orientation dependence of
the term $\boldsymbol{\nabla}_\mathbf{k}\Theta_\mathbf{k}\cdot \boldsymbol{\mathcal{E}}$, which 
in its turn depends on the polarization angle, $\theta$, of the incident field. Note that for low frequencies
(left panels, $\omega_p/t_\text{TB}=0.1$) only momentum values near the Dirac points are excited. On the other hand, for $\omega_p/t_\text{TB}=2$ the brightest spots occur at the $\mathbf{M}-$point. Also note
that the $\mathbf{M}-$points are not all excited at the same time, but depend on the polarization of the 
pumping field. This result constrasts with the case where the frequency of the pumping field pumps electrons to the Dirac cone (left panels). In this case, all the Dirac points are excited simultaneously.}
   \label{distfig}
\end{figure*}

\subsection{Steady-State Solution}

As shown in Fig. \ref{rtime}, the electronic distribution converges to a well defined value which we 
calculate in Appendix \ref{apdem}, when the electric field is given by a monochromatic wave with pumping frequency $\omega_p$ and intensity ${\mathcal E}_0$: $\boldsymbol{\mathcal E}=\frac{1}{2}\boldsymbol{\mathcal E}_0 e^{i\omega_p t} + \text{h.c.}$. In this case the  steady-state solution for the densities can be written as:
\begin{subequations}
\bea
n_{c,\mathbf{k}} = \frac{(1+\alpha_\mathbf{k})f_{c,\mathbf{k}}+ \alpha_\mathbf{k} f_{v,\mathbf{k}} }{1+2\alpha_\mathbf{k}} ,
\\
n_{v,\mathbf{k}} = \frac{(1+\alpha_\mathbf{k})f_{v,\mathbf{k}} + \alpha_\mathbf{k} f_ {c,\mathbf{k}}}{1+2\alpha_\mathbf{k}} ,
\eea \label{nrho}
\end{subequations}
and from Eq. (\ref{rhodef})
\be
\rho_{\mathbf{k}}=\frac{\alpha_\mathbf{k}}{1+2\alpha_\mathbf{k}}\left( f_{v,\mathbf{k}} - f_ {c,\mathbf{k}}\right), \label{rhoeq}
\ee
where $\alpha_\mathbf{k}$ is given by Eq. (\ref{alphak}) and $\rho_\mathbf{k}$ represents the deviation from the equilibrium Fermi-Dirac distribution. It should be noted that the steady state distribution functions are not given 
by a Fermi-Dirac distribution, but can be written in terms of combinations of $f_{c,\mathbf{k}}$
and $f_{v,\mathbf{k}}$.
The non-equilibrium distribution depends on the pumping frequency $\omega_p$ and on the complex electric field $\boldsymbol{\mathcal{E}}_0$. For the linear
polarization the electric field can be written as a real quantity that depends on the intensity of the electric field ${\mathcal E}_0$ and the angle of 
polarization $\theta$. 

The distribution $\rho_\mathbf{k}$ is plotted in Fig. \ref{distfig} for different pumping frequencies $\omega_p$ and polarization angles $\theta$ of the pumping field. As the pumping frequency increases, the electronic distribution departs from the Dirac points (the corners of the blue hexagon). At $\hbar\omega_p=2t_\text{TB}$, the $\mathbf{M}$-point is populated. For $\hbar\omega_p>2 t_\text{TB}$, the electronic distribution becomes a circle around the $\Gamma$-point  (center of the hexagon). We also  see
in this figure the polarization dependence coming from the term $\boldsymbol{\nabla}_\mathbf{k}\Theta_\mathbf{k}\cdot \boldsymbol{\mathcal{E}}$. For example, although we have three independent $\mathbf{M}$-points, for $\hbar\omega_p=2t_\text{TB}$ and $\theta=\pi/2$ only two are populated. We can use Fig. \ref{quiver} to predict, for a given pumping polarization, what points in the Brillouin zone can be optically populated, noting that the electric
field $\boldsymbol{\mathcal{E}}$ and the vector field $\boldsymbol{\nabla}_\mathbf{k}\Theta_\mathbf{k}$ need to be parallel to maximize the electronic occupation. This anisotropy in the population
of the $\mathbf{M}$-points is at the heart of other anisotropic effects that we will discuss ahead.

\section{Intraband transitions of the non-equilibrium gas due to the probe field} \label{itfpe}

The optical response of graphene is determined by intraband and interband transitions\cite{Peres2010}. As shown in the previous section,  in the steady state the pumping field changes the electronic distribution and therefore the 
optical conductivity of the material. This quantity is related to the charge-charge correlation function
of graphene. The charge-charge correlation function can be calculated  using the new electronic distribution obtained in Eqs. (\ref{nrho}) and (\ref{rhoeq}), instead of the equilibrium Fermi-Dirac distribution, as:
\be
\chi(\mathbf{q},\omega) =  \frac{2e}{\hbar a_0^2}\sum_{\lambda,\lambda^\prime=\pm} \int_{1^{\circ}\text{BZ}} \frac{d^2\mathbf{k}}{(2\pi)^2} \frac{n^\lambda_{\mathbf{k+q}}-n^{\lambda^\prime}_{\mathbf{k}}}{\omega- \omega_{\mathbf{k,q}}^{\lambda,\lambda^\prime}+i\varepsilon}  N_{\lambda^\prime\lambda}^{\mathbf{k},\mathbf{q}},
\ee
where the factor $2$ accounts for the spin degeneracy, $n^\lambda_{\mathbf{k}}$ is the electronic distribution given by Eq. (\ref{nrho}), $\hbar\omega_{\mathbf{k,q}}^{\lambda^\prime,\lambda}=\lambda^\prime E_\mathbf{k}-\lambda E_\mathbf{k+q}$ is the energy transition, and the overlap of the eigenfunctions is given by:
\be
 N_{\lambda^\prime\lambda}^{\mathbf{k},\mathbf{q}}=\frac{1}{2}\left[1+\lambda^\prime\lambda\cos\left(\Theta_\mathbf{k}-\Theta_\mathbf{k+q} \right) \right], \label{nck}
\ee

Using Eq. (\ref{rhodef}) to split the density into the equilibrium $f^{c/v}_\mathbf{k}$ and fluctuation $\rho_\mathbf{k}$ parts, the susceptibility can be decomposed into an equilibrium $\chi_0(\mathbf{q},\omega)$ part, that is calculated using the equilibrium Fermi-Dirac distribution, and two pumped components, one intraband and the other  interband, as:
\be
\chi(\mathbf{q},\omega) = \chi_0(\mathbf{q},\omega)+\chi^\text{intra}_\text{pump}(\mathbf{q},\omega)+ \chi^\text{inter}_\text{pump}(\mathbf{q},\omega), \label{split}
\ee
where $\omega$ is the frequency of the probe. The intraband pumped component of the susceptibility reads
\be
\chi^\text{intra}_\text{pump}(\mathbf{q},\omega)=\frac{2e}{\hbar a_0^2}\sum_{\lambda=\pm}\int_{1^{\circ}\text{BZ}}  \frac{d^2\mathbf{k}}{(2\pi)^2} \frac{ \lambda\left(\rho_\mathbf{k+q} -\rho_\mathbf{k}\right)}{\omega- \lambda \omega_{\mathbf{k,q}}+i\varepsilon}  N_{\lambda,\lambda}^{\mathbf{k},\mathbf{q}}, \label{eqi}
\ee
where $\omega_{\mathbf{k,q}}=E_\mathbf{k}-E_\mathbf{k+q}$.
Note that $\chi^\text{intra}_\text{pump}$ is determined by the deviations to the Fermi-Dirac distribution. 
Since we are only interested in the physics of the electron gas created in the conduction band, we neglect in the 
following the contribution from the interband transitions $\chi^\text{inter}_\text{pump}(\mathbf{q},\omega)$  to the total susceptibility, an assumption that is valid when $\omega_p\gg\omega$. This is always the case
in this work as we are considering pumping to the $\mathbf{M}$-point, which resides in the UV-part of the electromagnetic spectrum.

We now want to introduce the effect of relaxation into the calculation of the charge-charge susceptibility.
This can be done using Mermin's approach developed for the 3D electron gas\cite{mermin1970}.
Following Mermin's work and making the necessary modifications for the graphene case,
the total susceptibility, taking into account relaxation processes, is given by:	
\be
\chi^M(\mathbf{q},\omega)= \frac{\left(1+i(\tau\omega)^{-1}\right)\chi(\mathbf{q},\omega+i\gamma)}{1+i(\tau\omega)^{-1} \chi(\mathbf{q},\omega+i\gamma)/\chi(\mathbf{q},\omega=0)}, \label{eqmerm}
\ee
where $1/\tau=\gamma$ is the relaxation rate. For calculating Mermin's susceptibility we need the Lindhard susceptibility $\chi(\mathbf{q},\omega+i\gamma)$, which
 needs to be computed for a complex frequency.  In addition we also need the static susceptibility $\chi(\mathbf{q},\omega=0)$.
The dielectric function can be obtained from the suceptibility as\cite{Goncalves2016}:
\be
\varepsilon(\mathbf{q},\omega)= 1-v_\mathbf{q}\chi^M(\mathbf{q},\omega), \label{dieexp}
\ee
where $v_\mathbf{q}=e a_0/(2\varepsilon_m q)$ is the 2D Fourier transform of the Couloumb potential and $\varepsilon_m=\frac{\varepsilon_1+\varepsilon_2}{2}$ is the effective dielectric constant of the environment for a graphene clad between two media of dielectric constants $\varepsilon_1$ and $\varepsilon_2$. We recall that the term $a_0$ appears in $v_\mathbf{q}$ because the wave number $q$ is measured in units of the inverse lattice parameter $a_0^{-1}$.
For consistency with the Mermin's formula, we take $\gamma_p=\gamma_0=\gamma$ 
in the forthcoming equations. In the all the figures we have kept $\gamma_p\ne\gamma_0$,
which is in agreement with Mermin's equation in the long wavelength limit.

\section{Long wavelength limit: anisotropic plasmon dispersion relation } \label{lwl}

The calculation of the integral in Eq. (\ref{eqi}) needs to be done for every different frequence $\omega$ and wavenumber $\mathbf{q}$. However, as shown
in Fig. \ref{thetadep}, as $\mathbf{q}$ decreases the conductivity reaches the long wavelength limit and
we can show that in this regime
 the susceptibility in Eq. (\ref{eqi}) behaves as $q^2$. In this limit, the static susceptibility  appearing in the denominator of Eq. (\ref{eqmerm}) tends to a constant value when $q\rightarrow0$, and therefore
 Eq. (\ref{eqmerm}) becomes:
\be
\chi^M(\mathbf{q},\omega)= \left(1+i(\tau\omega)^{-1}\right)\chi(\mathbf{q},\omega+i\gamma). \label{eqmerm2}
\ee
We now split the susceptibility in the right hand side of the Eq. (\ref{eqmerm2}) in the same way as we did in Eq. (\ref{split}) ---that is in an equilibrium and an out-of-equilibrium parts. The equilibrium component $\chi^0(\mathbf{q},\omega)$ can be approximated by the Drude term for $\hbar\omega<2 E_F$, where $E_F$ is the Fermi energy:
\be
\chi^\text{doped}(\mathbf{q},\omega)= \frac{4eE_F}{\pi\hbar^2 a_0^2}\frac{q^2}{(\omega+i\gamma)^2}\,.\label{efsu}
\ee
For undoped graphene we have $E_F=0$ and the Drude contribution vanishes.
For the out-of-equilibrium component,  we obtain a similar expression in the long wavelength limit for the pumped susceptibility using Eq. (\ref{eqi}) (details of the calculations are given in Appendix \ref{saf}) in the form:
\be
\chi^\text{intra}_\text{pump}(\mathbf{q},\omega)=\sum_{i,j}C_{ij}\frac{q_i q_j}{(\omega+i\gamma)^2}\,. \label{chip3}
\ee
The term in the right hand side of Eq. (\ref{chip3}) corresponds to the intraband pumped contribution and can also be written as a quadratic dependence on the modulus of the wavevector $\mathbf{q}$.
This is one of the central results of this paper with far reaching implications. 

Comparing Eq. (\ref{chip3}) with the susceptibility of doped graphene in the long wavelength limit in Eq. (\ref{efsu}),  we can define an effective Fermi energy, that depends on the polarization angle $\varphi$ of the probe field relative to the graphene lattice,
as:
\be
E_F^\text{eff}(\varphi)=E_F+f_0+f_m\cos(2\varphi+\phi), \label{eff}
\ee
where $f_0$, $f_m$, and $\phi$ depend only on the properties of the pumping field ---$E_\text{pump}$, $\theta$,  and $\omega_p$--- which are defined in Appendix \ref{saf}.
Finally the susceptibility in Eq. (\ref{eqmerm2}) can be written as:
\be
\chi^M(\varphi,\omega)= \frac{4e}{\pi\hbar^2 a_0^2}\frac{E_F^\text{eff}(\varphi)q^2}{\omega(\omega+i\gamma)}\,.
\ee
The plasmon dispersion is obtained from the condition $\varepsilon(\mathbf{q},\omega)=0$ in Eq. (\ref{dieexp}), leading to:
\be
\hbar\omega(\varphi,q)=\sqrt{2 \alpha \frac{\hbar c}{a_0} E_F^\text{eff}(\varphi) q}-i\frac{\gamma}{2}, \label{pdisp}
\ee
where $\alpha\approx1/137$ is the fine structure constant of atomic physics. Equation (\ref{pdisp}) has the same $\sqrt{q}$ dependence as that of plasmons in doped graphene without the pumping field\cite{Jablan2009,Goncalves2016}. The difference lies in the presence of an effective Fermi energy 
$E_F^\text{eff}(\varphi)$ that depends on the direction of the wavevector. Equation (\ref{pdisp}) and is one of the central results of this work. Note that the dispersion will be anisotropic, as the effective Fermi energy depends
on the orientation of the pumping electric field relatively to the graphene lattice. Furthermore, even in the case of neutral graphene, the system support plasmons since $E_F^\text{eff}(\varphi)$ is finite
even for $E_F=0$, due to the constant illumination of the pumping field.

\section{The anisotropic conductivity of graphene under pumping} \label{tacgup}
In this section we show that in an out-of-equilibrium situation we can define an anisotropic optical conductivity for graphene. 
The optical conductivity tensor $\sigma_{ij}(\mathbf{q},\omega)$ can be obtained via the continuity 
equation:
\be
\mathbf{q}\cdot\mathbf{J}-\omega \rho=0, \label{conteq}
\ee
where $\rho$ is the charge density and $\mathbf{J}$ the surface density current. The
current is described by 
\be 
J_i=\sum_j \sigma_{ij}(\mathbf{q},\omega){\cal E}_j=i\sum_j \sigma_{ij}(\mathbf{q},\omega)q_j \Phi\,.
\ee
The previous result follows from the relation between the electric potential $\Phi$, with
well defined momentum $\mathbf{q}$, and the electric field $\boldsymbol{\mathcal{E}}$ via
the relation $\boldsymbol{\mathcal{E}}=-\boldsymbol{\nabla}\Phi=i\mathbf{q}\Phi$. On the other hand,
the charge density is obtained from the charge-charge susceptibility via $\rho=\chi^M(\mathbf{q},\omega) \Phi$. Thus, using Eq. (\ref{conteq}),
the relation between the conductivity tensor and the susceptibility is:
\be
\sum_{i,j} \sigma_{ij}(\mathbf{q},\omega) q_i q_j = i \omega \chi^M(\mathbf{q},\omega). \label{siseq}
\ee
The Equation (\ref{siseq}) is not enough to determine the conductivity tensor from the susceptibility, but 
in the long wavelength limit, $q\rightarrow0$, the dependence of each element of the conductivity tensor on the wavenumber 
disappear, and we can obtain three independent equations to the four quantities $\sigma_{ij}$. These three equations can be obtained changing the direction of the wavevector $\mathbf{q}$ or,  equivalently, we can compare the Taylor expansion of $\chi^M(\mathbf{q},\omega)$ to the left hand side of Eq. (\ref{siseq}). This procedure would give four equations but one of them would not be independent of the other three. The
missing equation can be obtained from the current-current response, calculated in appendix \ref{coc}, where
the intraband contributions to the conductivity tensor read
\be
\sigma_{ij}^\text{intra}(\mathbf{q},\omega) =\frac{2ie^2}{\hbar \omega S} \sum_{\mathbf{k}, \lambda=\pm} \frac{n^\lambda_\mathbf{k+\mathbf{q}/2}-n^\lambda	_{\mathbf{k}-\mathbf{q}/2}}{\omega-\lambda \omega^\text{intra}_{\mathbf{k},\mathbf{q}}+i\gamma} {\boldsymbol{v}_i}^\text{intra}_{\mathbf{k},\mathbf{q}} {\boldsymbol{v}_j}^\text{intra}_{\mathbf{k},-\mathbf{q}}, \label{ccresp}
\ee
with ${\boldsymbol{v}_i}^\text{intra}_{\mathbf{k},\mathbf{q}}$ defined in Appendix \ref{coc} and $S=N_c a_0^2$, where $N_c$ is the number of unit cells.
In this appendix an expression
for  the interband term is also provided. From Eq. (\ref{ccresp}) we can show that in the limit $\mathbf{q}\rightarrow0$, we have:
\be
\sigma_{ij}^\text{intra}(\mathbf{q}\rightarrow0,\omega)=\sigma_{ji}^\text{intra}(\mathbf{q}\rightarrow0,\omega),
\ee
thus it follows  from Eq. (\ref{siseq}) that:
\bea
\sigma^\text{pumped}_{ij}= \sigma_0\frac{C_{ij}}{\omega+i\gamma},
\eea 
where $\sigma_0=e^2/(4\hbar)$
and $C_{ij}$ are the coefficients of the expansion of $\chi^\text{pumped}(\mathbf{q},\omega)$ defined in Eq. (\ref{chip3}), and the total intraband conductivity can be written as function of an effective Fermi energy tensor $E_{ij}^\text{eff}$ as:
\be
\frac{\sigma_{ij}}{\sigma_0}=i\frac{4}{\hbar\pi} \frac{E_{ij}^\text{eff}}{\omega+i\gamma}, \label{sigef}
\ee
where 
we have defined the effective Fermi energy tensor as:
\be
E_{ij}^\text{eff}=E_F\delta_{ij}+\frac{\pi}{4}C_{ij}.
\ee
 Although the tensor $E_{ij}^\text{eff}$ can be reduced to diagonal form by a rotation, doing so we loose the direct connection of the tensor components to the orientation of the graphene lattice.

We can also  define the longitudinal conductivity along the direction defined by the unit vector $\mathbf{u}_\varphi$ for
a probing electric field of the form $\boldsymbol{\mathcal{E}}=\mathcal{E}_0\mathbf{u}_\varphi$ as:
\be
\sigma_\mathbf{\varphi}= \frac{\mathbf{J}\cdot \mathbf{u}_\varphi}{\mathcal{E}_0}=\frac{4i}{\hbar\pi}\frac{E_F^\text{eff}(\varphi)}{\omega+i\gamma}, \label{clv}
\ee
where the angle $\varphi$ (the polarization angle of the probing field) is the same as that of the momentum $\mathbf{q}$, since 
the electric field is proportional to $\mathbf{q}$ via the gradient of the potential.

It is worth remembering that the effective parameters $f_0$,$f_m$, and $\phi$ depend solely on the
pumping field properties, that is, on the intensity $\mathcal{E}_0$, the polarization angle $\theta$, and the frequency $\omega_p$.

\section{Numerical results} \label{nr}
  
As shown before (see Fig. \ref{distfig}), graphene under intense and energetic light pumping presents a strong anisotropic electronic distribution. This changes the
optical response due to intraband and interband transitions. In doped graphene, without electromagnetic pumping, the intraband transitions
dominates for photon energy $\hbar\omega<2E_F$, while interband transitions dominate for $\hbar\omega>2E_F$ \cite{Peres2010}. For 
pumped graphene, we have a similar result, where intraband transitions dominate for $\omega<\omega_p$, where $\omega_p$ is the frequency
of the pumping radiation, and interband transitions dominates for $\omega\approx\omega_p$. 

To show the effects of the pumping in graphene, we solve numerically the Eq. (\ref{eqi}) and compute the pumped component of the 
intra-band susceptibility. The $\chi_0(\mathbf{q},\omega)$ component is calculated with the analytical expressions derived with the Dirac equation \cite{Goncalves2016},
since it is not necessary here to account for the full band structure of graphene.
This is because for the equilibrium distribution, the probe frequency $\omega$ in the range we are considering can only excite electron-hole pairs around the Dirac cone. This is not the case for the pumped electron gas around the $\mathbf{M}$-point, that cannot be described by the Dirac equation. The out-of-equilibrium distribution is calculated using Eq. (\ref{rhoeq}).

\begin{figure}  
   \vspace{0.5cm}
   \centering
   \includegraphics[scale=0.24]{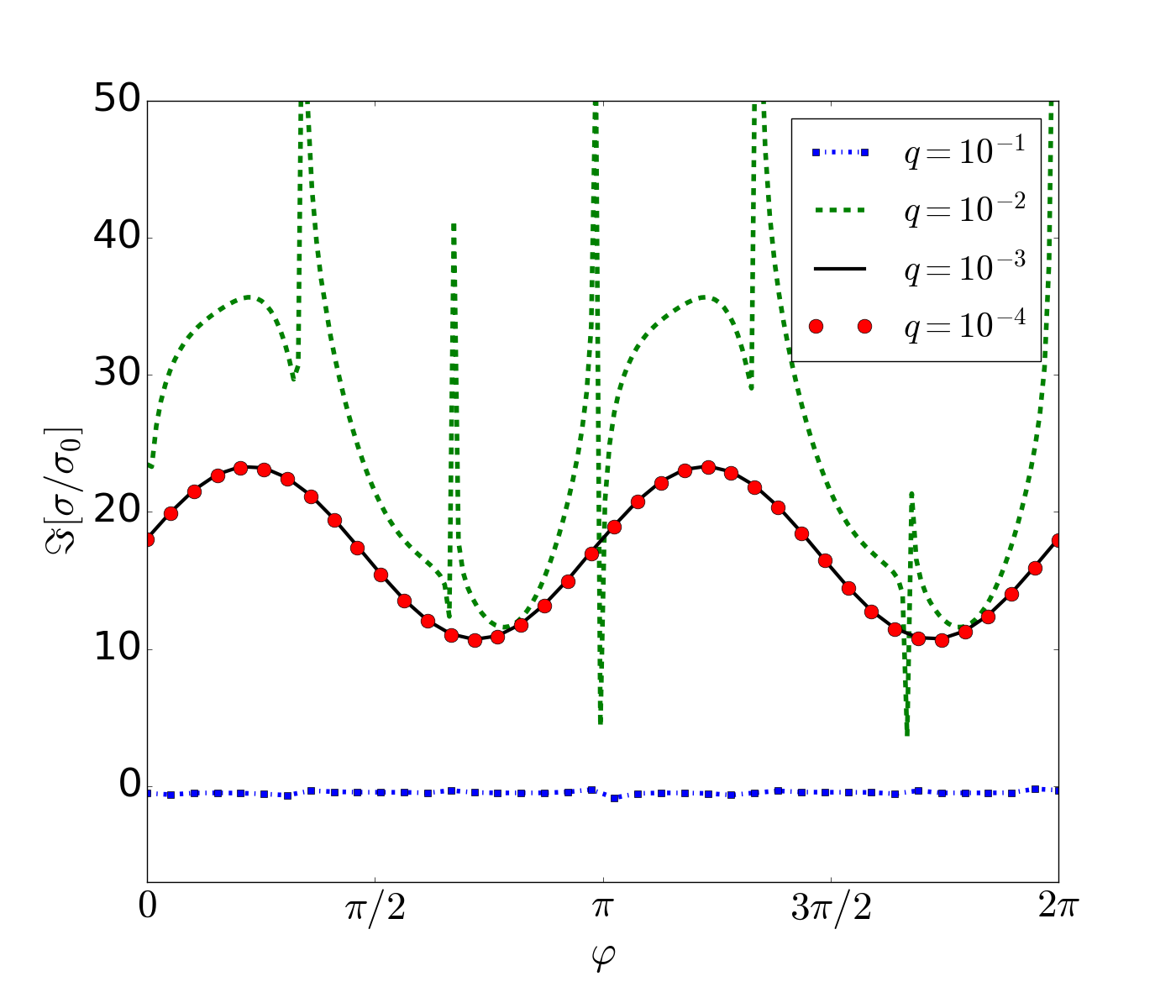}
   \caption{(Color on-line.) Imaginary part of the longitudinal conductivity as function of the polarization angle of the pump field for different values of $q$. The parameters are: $\mathcal{E}_0=0.707$ GV/m, $\hbar\omega_p=2t_\text{TB}$, $\theta=\pi/4$, $E_F=0.2$ eV and $\hbar\gamma_0=14$ meV, and $\hbar\gamma_p=28$ meV. For $q<10^{-3}$ the susceptibility reaches the long wavelength limit. For $q=10^{-2}$ we can see the strong influence of the static susceptibility (see Fig. \ref{static0}).} \label{thetadep}
\end{figure}

We plot in Fig. \ref{thetadep}  the imaginary part of the longitudinal conductivity, Eq. (\ref{clv}), as function of the probe incidence angle $\varphi$, 
for different wave numbers $q$; this quantity controls the dispersion of the surface plasmon-polariton in the out-of-equilibrium electron gas, as will be discussed in a forthcoming section. We show that the long wavelength limit is reached around $q=10^{-3}$, where the conductivity have a co-sinusoidal shape as predicted by Eq. (\ref{eff}). Note that $q$ is measured in units of $1/a_0$. In the same figure we also
depict an example of the longitudinal conductivity away from the long wavelength limit
($q=10^{-2}$). It is clear that in this regime the distribution is, for some $\varphi$
values, substantially different form the analytical approximation. Since we are interested here in the long wavelength limit, this results does not interest us and will not be discussed
further.

\begin{figure} 
   \vspace{0.5cm}
   \centering
   \includegraphics[scale=0.24]{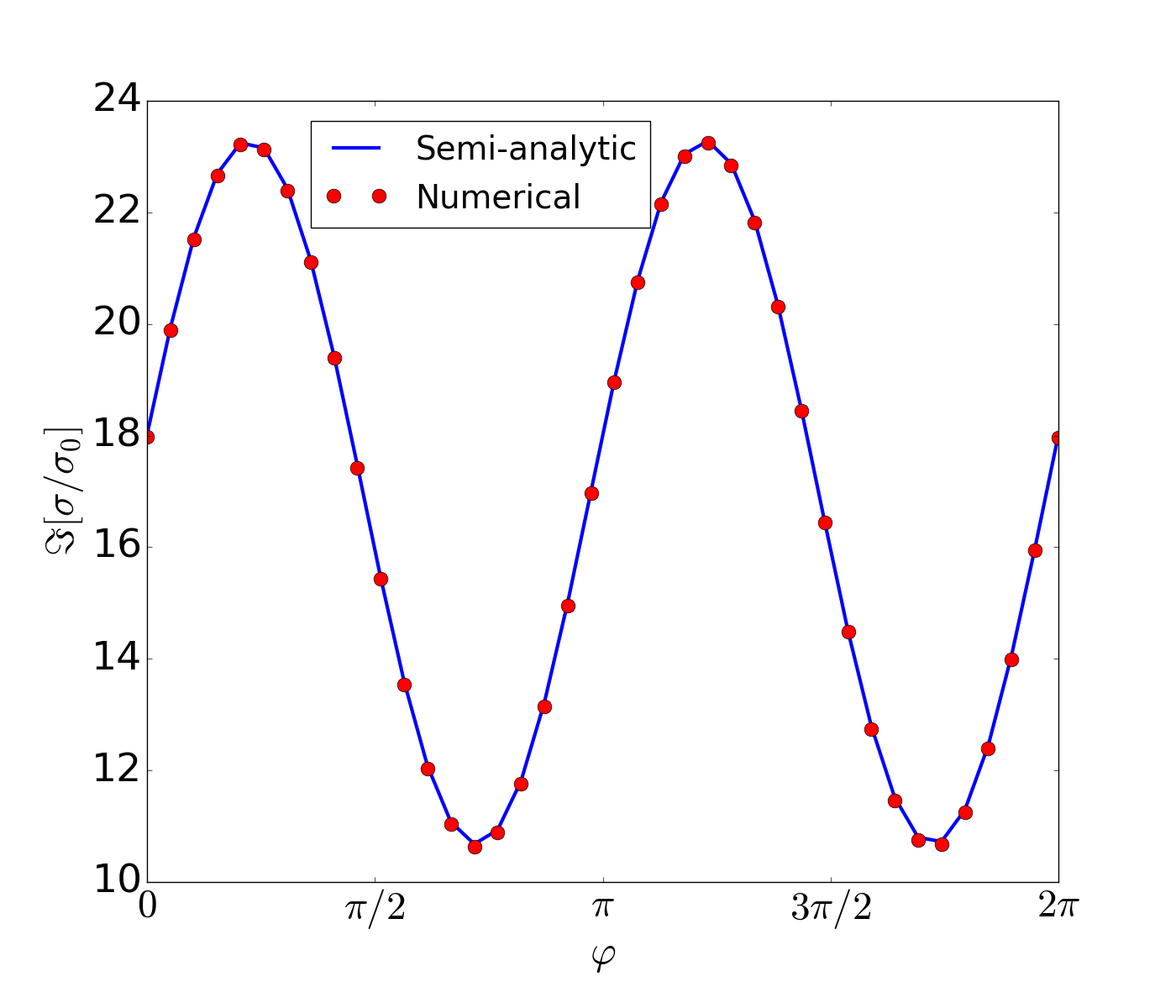}
   \caption{(Color on-line.) Comparision between the semi-analtical approach and the numerical one for the imaginary part of the longitudinal conductivity, showing the validity of the semi-analytical approximation obtained in section \ref{lwl}.   
The dots correspond to the black solid curve in Fig. \ref{thetadep} and the 
solid line is the semi-analytical calculation.   Note that $\Im\sigma\in[10\sigma_0,24\sigma_0]$; the difference between the maximum and the minimum of the conductivity depends on the magnitude of $\mathcal{E}_0$. 
The parameters are: $q=10^{-3}$, $\mathcal{E}_0=0.707$ GV/m, $\hbar\omega_p=2t_\text{TB}$, $\theta=\pi/4$, $E_F=0.2$ eV, and $\hbar\gamma_0=14$ meV, and $\gamma_p=28$ meV. } \label{thetacomp}
\end{figure}

Figure \ref{thetacomp} shows the numerically computed imaginary part of the longitudinal  conductivity, as function of the probing polarization angle $\varphi$, for $q=10^{-3}$, as defined by Eq. (\ref{clv}),  compared  with the semi-analytical result, which depends on the  effective Fermi energy defined in Eq. (\ref{eff}). The two approaches show a very good agreement, showing that indeed for $q=10^{-3}$ the system is already
in the long wavelength regime. The oscillatory variation of 
the imaginary part of the longitudinal  conductivity
will lead to an anisotropy in the spectrum  of the
surface plasmon-polariton, as it is this quantity that determines
the behavior of the latter. Note that $\Im\sigma\in[10\sigma_0,24\sigma_0]$ (see Fig. \ref{thetacomp}).

From here on our analysis is focused on two ways 
of parameterizing the effective Fermi energy. In the first approach, we discuss Eq. (\ref{eff}), which is suitable for analyzing the susceptibility (\ref{efsu}) and plasmons modes (\ref{pdisp}) at long wavelengths. In the second approach, the result of Eq. (\ref{sigef}) is useful to calculate the optical conductivity at long wavelengths and the dispersion of the surface plasmon-polariton, defined by Eq. (\ref{sppeq}).

\begin{figure}  
   \vspace{0.5cm}
   \centering
   \includegraphics[scale=0.34]{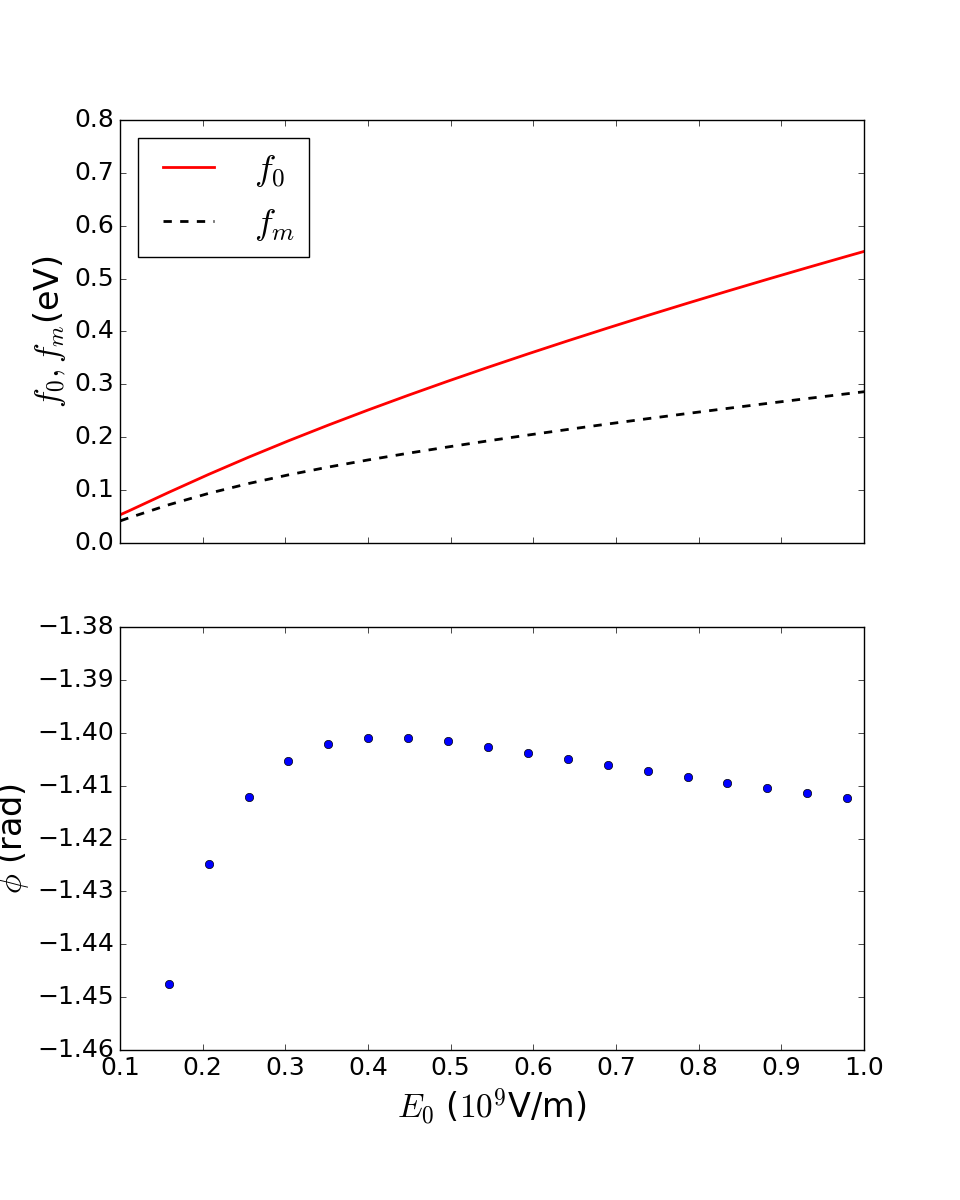}
   \caption{(Color on-line.) 
Dependence of the parameters $f_0$, $f_m$, and $\phi$ on the  
   intensity of the pumping radiation.  The parameters are: $\theta=\pi/4$, $\hbar\omega_p=2t_\text{TB}$, $E_F=0.2$ eV, $\hbar\gamma_0=14$ meV, and $\hbar\gamma_p=28$ meV.  The importance of the parameters
   $f_0$ snd $f_m$ grows with the intensity of the pumping field.
   The minimum value of field intensity considered in this figure is 0.1 GV/m. We see that the anisotropy
   is observable for this field intensity.
We note that the field intensities scanned in this figure are experimentally attainable.} \label{intensity}
\end{figure}

Figure  \ref{intensity} shows that the parameters $f_0$ and $f_m$ have a strong dependency on the intensity of the pumping field. The angle $\phi$, in contrast, changes very little by  as much as $\sim0.1$ rad, and tends to saturate 
for large intensity fields. The  parameters  $f_0$ and $f_m$ can have a strong impact in the 
optical response of the system, depending on the initial doping level of graphene, characterized by $E_F$. 
For large doping, the effect of $f_0$ and $f_m$ is small, except for large pumping field intensities. However, for vanishing small Fermi energies, the effect of these two 
parameters have a large impact in the optical properties of the system, as the effective Fermi energy is essentially controlled by them.

\begin{figure} 
   \vspace{0.5cm}
   \centering
   \includegraphics[scale=0.34]{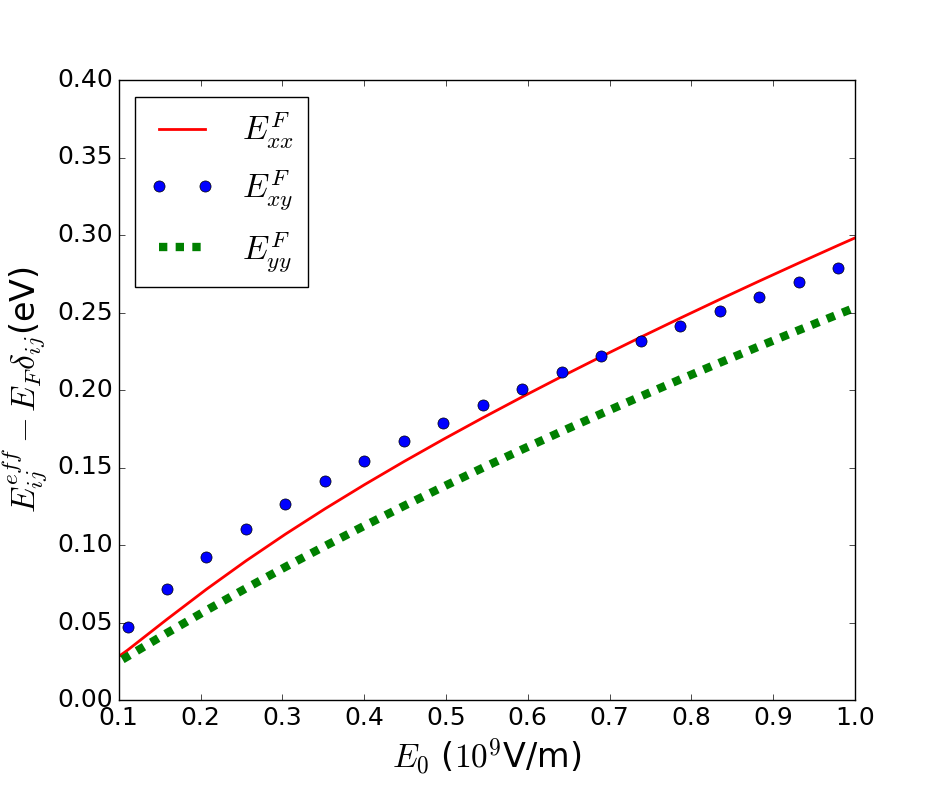}
   \caption{(Color on-line.) 
   Dependence of the effective Fermi energy $E^\text{eff}_{ij}-E_F\delta_{ij}$ on the  
   intensity of the pumping radiation (in GV/m). 
The anisotropy grows with the increase of $\mathcal{E}_0$.   
   The parameters are: $\theta=\pi/4$, $\hbar\omega_p=2t_\text{TB}$, $E_F=0.2$ eV, $\hbar\gamma_0=14$ meV, and $\hbar\gamma_p=28$ meV. 
The minimum value of field intensity considered in this figure is 0.1 GV/m.   
   \label{fig_EF_E0}    
   }
\end{figure}

\begin{figure}  
   \vspace{0.5cm}
   \centering
   \includegraphics[scale=0.4]{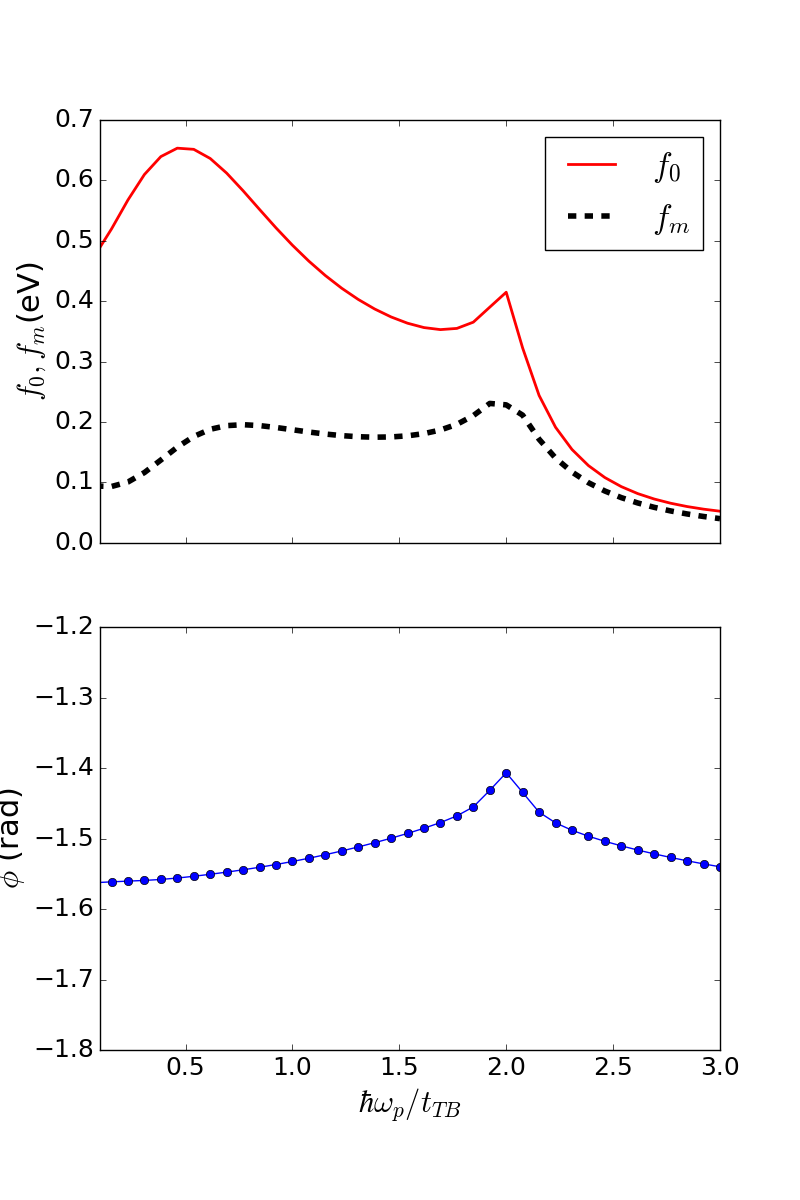}
   \caption{(Color on-line.) Dependence of the parameters $f_0$, $f_m$, and $\phi$ on the energy of the photon of the pumping field. The parameters are: $\mathcal{E}_0=0.707$ GV/m, $\theta=\pi/4$, $\hbar\omega_p=2t_\text{TB}$, $E_F=0.2$ eV, $\hbar\gamma_0=14$ meV, and $\hbar\gamma_p=28$ meV. A non-monotonous dependence on $\omega_p$ is seen for the three parameters.   We must however stress that for 
$\hbar\omega_p\approx 0$ the behavior of  $f_0$, $f_m$, and $\phi$ is
not accurate, as we have not included the effect of interband transitions,
due to probe of frequency  $\omega$, which become 
relevant for $\omega_p\sim\omega$, specially in the case of neutral graphene.  See Fig. \ref{fig_EF_omegapump} for a discussion of the 
position of the  maximum of $f_0$ and $f_m$ located at low energies.
    \label{pumpd}}
\end{figure}

\begin{figure}  
   \vspace{0.5cm}
   \centering
   \includegraphics[scale=0.26]{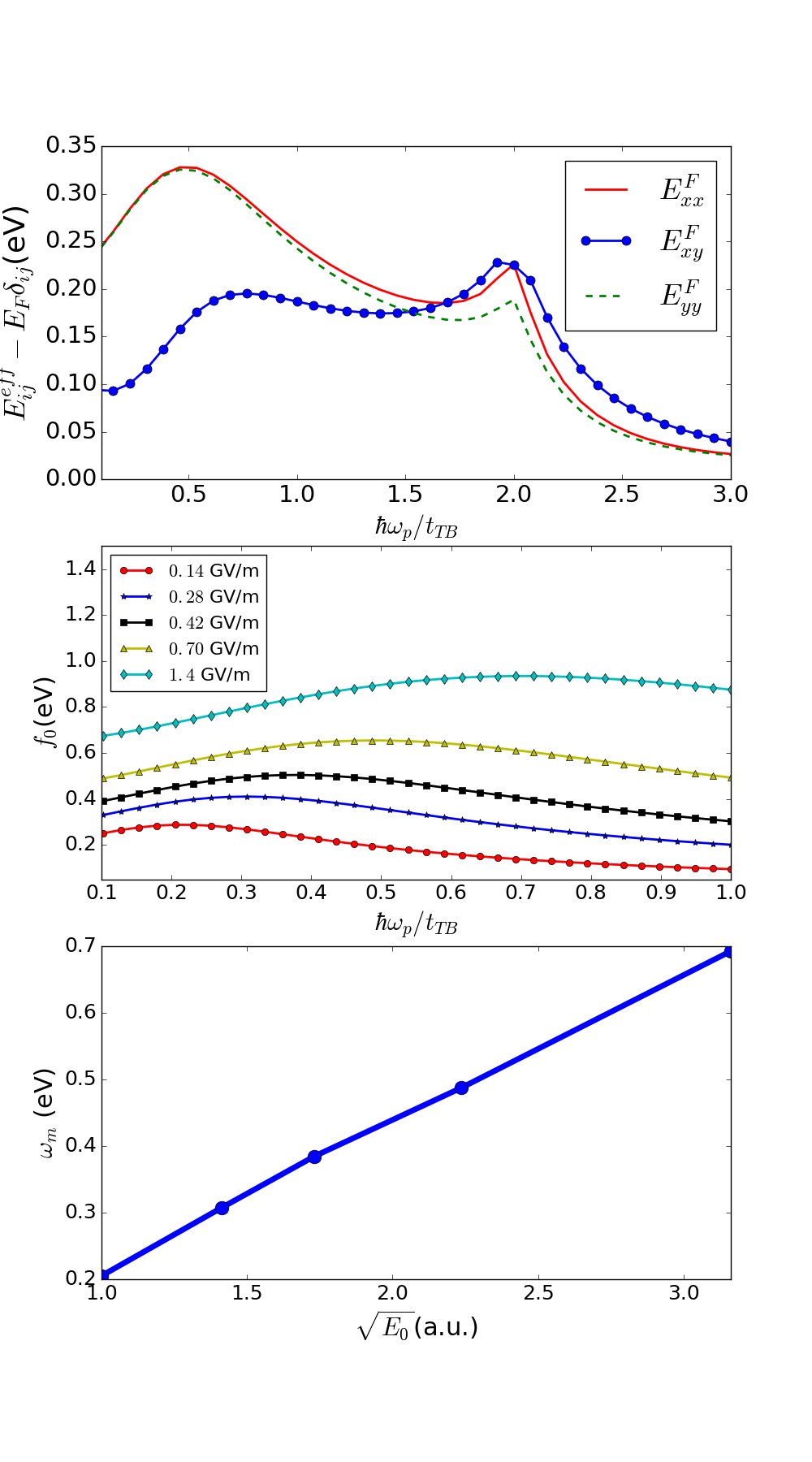}
   \caption{(Color on-line.) Top panel: Dependence of the components of the 
   effective Fermi energy tensor $E^\text{eff}_{ij}-E_F\delta_{ij}$ on the pumping field frequency. The parameters are: $\mathcal{E}_0=0.707$ GV/m, $\theta=\pi/4$, $E_F=0.2$ eV, $\hbar\gamma_0=14$ meV, and $\hbar\gamma_p=28$ meV. 
Note the local maximum of the Fermi-energy tensor-elements around
the photon energy $\hbar\omega=2t_\text{TB}$. 
Also note that the largest difference between $E^F_{xx}$ and $E^F_{yy}$ occurs at the $\mathbf{M}-$point which implies the largest anisotropy
in the properties of the system, including the plasmon spectrum.
We must stress that for 
$\hbar\omega_p\approx 0$ the behavior the effective Fermi energy components are
not accurate, as we have not included the effect of interband transitions,
due to the probe of frequency $\omega$, which become 
relevant for $\omega_p\sim\omega$, specially in the case of neutral graphene.   Central panel: Zoom in of the dependence of the parameter
$f_0$ with $\mathcal{E}_0$ near the absolute maximum.
Bottom panel:
Scaling of the frequency of the maximum, $\omega_m$, with the 
$\sqrt{\mathcal{E}_0}$ (right panel); the linear scaling is evident.
The values of $\omega_m$ are extracted from the central panel,
and correspond to the position of the maximum of the curves for
$f_0$. Note that the larger $\mathcal{E}_0$ is the broader is the maximum
and more intense is $f_0$.
   \label{fig_EF_omegapump}}
\end{figure}

The dependence of the effective Fermi energy $E^F_{ij}$ on the  
   intensity of the pumping radiation is depicted in Fig. \ref{fig_EF_E0}. A clear anisotropy is seen in this quantity.
   Particularly interesting is the finite value of $E^F_{xy}$, which leads to a finite off-diagonal term for the 
   non-equilibrium optical conductivity. The dependence of the parameters
   $f_0$, $f_m$, and $\phi$ on the energy of the pumping photons is depicted in Fig. 
   \ref{pumpd}. We see that there is a non-monotonous dependence on $\omega_p$ with
   a local maximum (for $f_0$, $f_m$, and $\phi$) when the photon energy
   is equal to the electronic transition at the $\mathbf{M}$-point ($\hbar\omega_p/t_\text{TB}=2$). This is, most likely, due to the enhanced
   density of states associated with the van-Hove singularity.
In Fig. \ref{fig_EF_omegapump} the effective Fermi energy is depicted as function of the frequency of the 
pumping field. Clearly its behavior is controlled by the values of the parameters $f_0$, $f_m$, and $\phi$,
as can be seen from comparing Figs. \ref{pumpd} and \ref{fig_EF_omegapump}. Again a local maximum
is seen at the value of photon energy given by $\hbar\omega=2t_\text{TB}$. 

It is worthwhile to remark that the absolute maximum of the $f_0$ and
$f_m$ parameters, in Fig. \ref{pumpd}, takes place for $\hbar\omega_p\approx0.5t_\text{TB}$
($\sim$ 1.4 eV), leading to an out-of-equilibrium gas with a larger effective
Fermi energy than when the system is pumped with photons of frequency
$\hbar\omega_p\sim2t_\text{TB}$. This energy scale is controlled 
by the electric field intensity. Indeed, the system has an energy scale $\Delta$,
for the parameters of Fig. \ref{pumpd}, given by
\begin{equation}
\Delta/t_\text{TB} \sim \sqrt{
\frac{{\cal E}_0a_0}{t_\text{TB}}}\sim0.2\,,
\end{equation}
which is of the same order of magnitude of $\hbar\omega_p\approx0.5t_\text{TB}$, the position of the absolute maximum of the parameters
$f_0$ and $f_m$.  
Note that apart from the gradient of the phase $\Theta_\mathbf{k}$,
${\cal E}_0a_0$ is essentially the Rabi frequency.
We have verified that by reducing the field intensity by five times, the position of the maximum red-shifts
to an energy of about two times smaller the value of $\hbar\omega=0.5t_\text{TB}$. This effect is represented in the bottom panel of Fig. \ref{fig_EF_omegapump}. The scaling of the position of the maximum of $f_0$ with $\sqrt{{\cal E}_0}$ is evident.  
Note, however, that, for these energy scales,  the anisotropy for the plasmon spectrum will be very
small, as $E^F_{xx}\approx E^F_{yy}$.
Let us also note here that
the intensity of the density of the states
at the van-Hove singularity is presumably controlled by the value of $\gamma_0$: the 
larger this parameter is the smaller is the density of states at the 
$\mathbf{M}-$point, which otherwise would be a divergence in the absence of 
relaxation.

\begin{figure}  
   \vspace{0.5cm}
   \centering
   \includegraphics[scale=0.35]{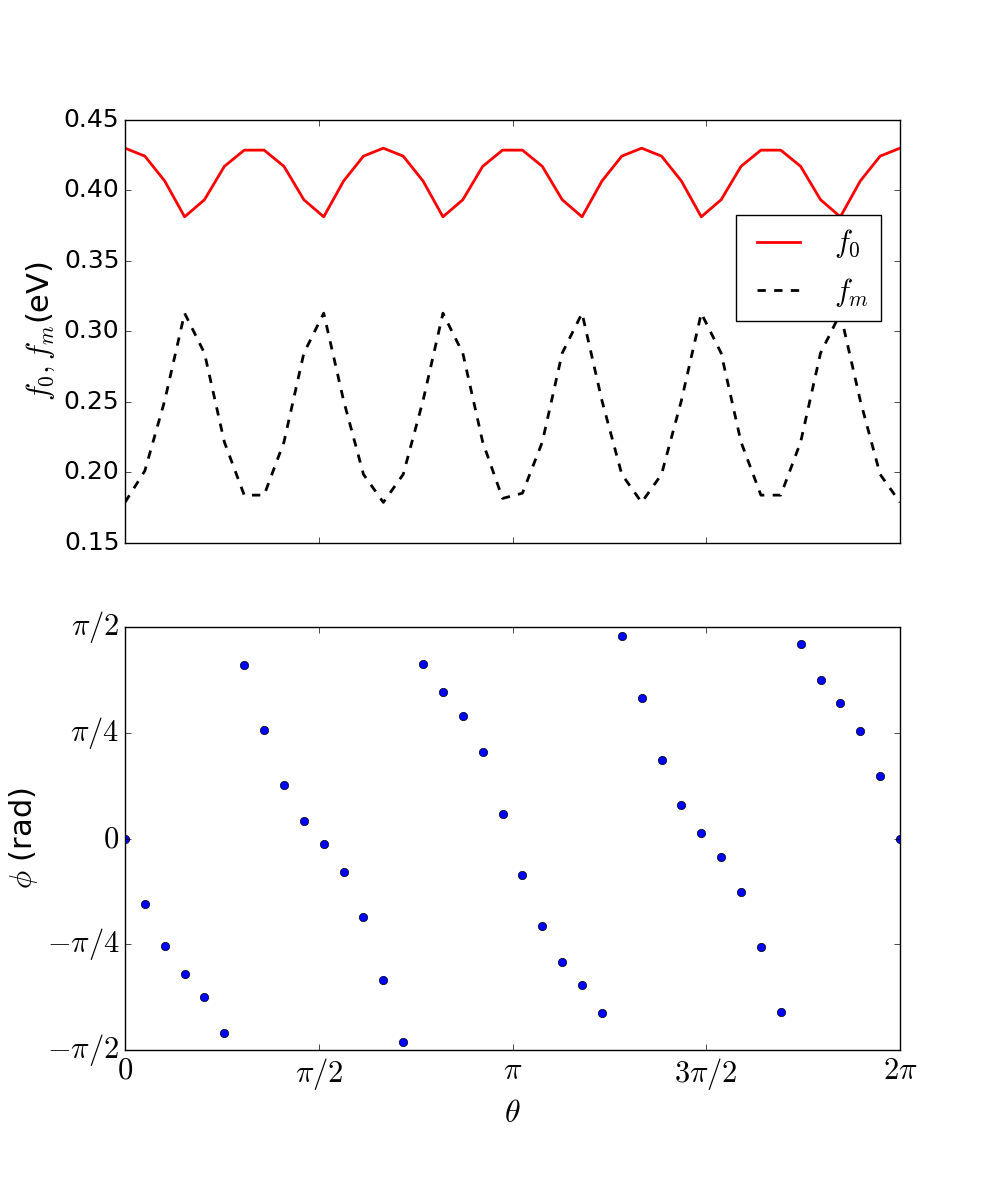}
   \caption{(Color on-line.) Dependence of the parameters determining the effective Fermi energy on the pumping
   polarization angle.  
Note that for some values of $\theta$ the magnitudes of $f_0$ and $f_m$ are almost 
identical. Also the angle $\phi$ varies substantially with $\theta$.   The largest anisotropy in the properties of the system occurs for the largest difference between $f_0$ and
$f_m$.  
   The parameters are: $\mathcal{E}_0=0.707$ GV/m, $\hbar\omega_p=2t_\text{TB}$, $E_F=0.2$ eV, $\hbar\gamma_0=14$ meV, and $\hbar\gamma_p=28$ meV. } \label{pangle}
\end{figure}

\begin{figure} 
   \vspace{0.5cm}
   \centering
   \includegraphics[scale=0.26]{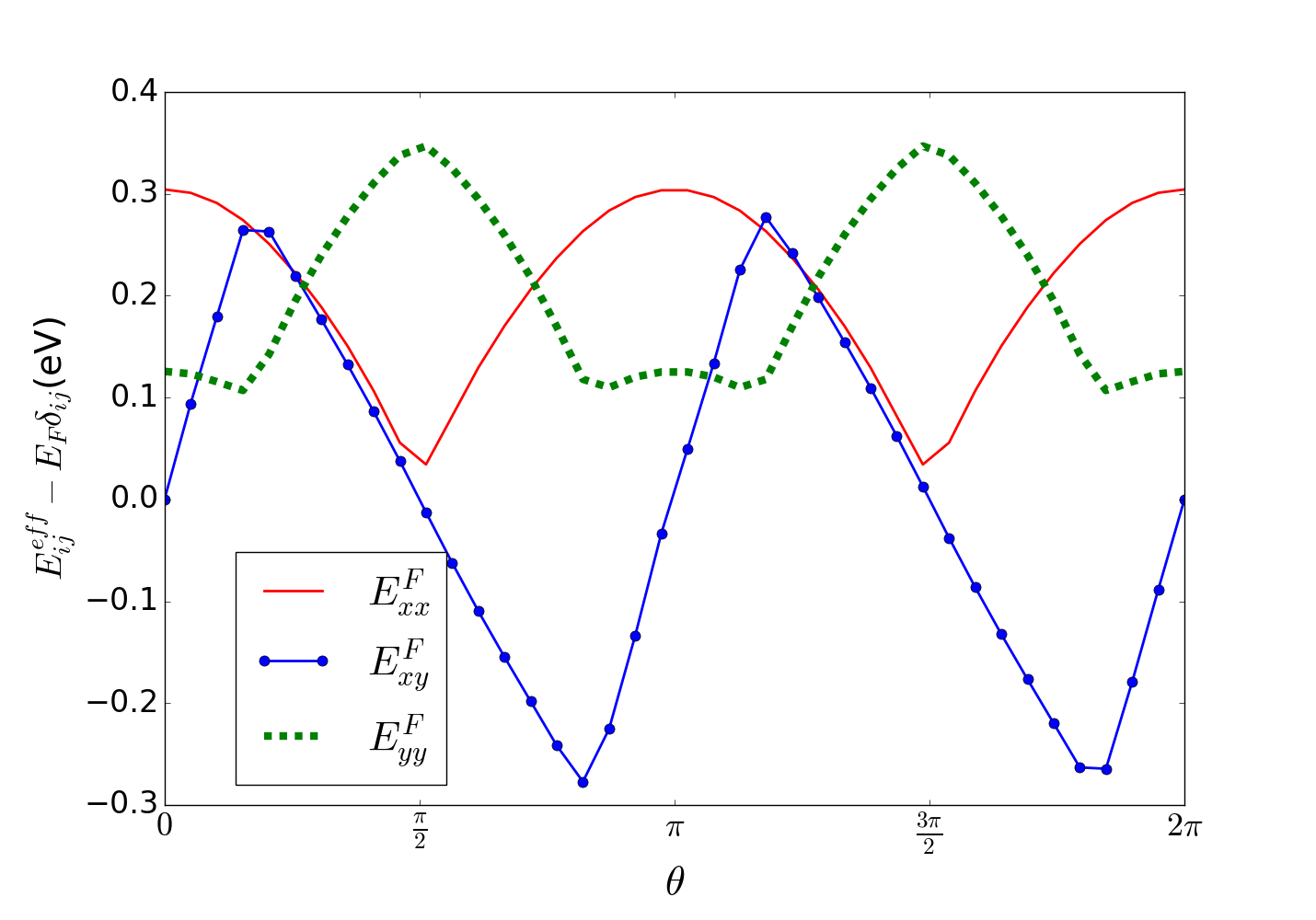}
   \caption{(Color on-line.) Dependence of the effective Fermi-energy $E^\text{eff}_{ij}-E_F\delta_{ij}$  on the polarization angle of the pumping field.  We emphasize that our calculations
   take the three $\mathbf{M}-$points into account simultaneously since we are making a tight-binding calculation. Therefore there is no cancellation of $E_{xy}$.
   The parameters are: 
  $\mathcal{E}_0=0.707$ GV/m, $\hbar\omega_p=2t_\text{TB}$, $E_F=0.2$ eV, $\hbar\gamma_0=14$ meV, and $\hbar\gamma_p=28$ meV. Note the periodic behavior of the different parameters.} \label{angsig}
\end{figure}

In Figs. \ref{pangle} and \ref{angsig} we show the strong anisotropy in the optical response. The parameter $f_m$, that measures the amplitude of the effective Fermi energy modulation,  
has maxima where the $f_0$ presents minima for some specific angles. This is the origin of the strong anisotropy
in the optical response of the system, which imparts in the anisotropy of the dispersion relation of the plasmons. In the Fig. \ref{pangle}  the parameter $\phi$ is also depicted
showing a strong variation with the angle of polarization of the pumping field.
The strong variation of $f_0$, $f_m$, and $\phi$ on $\theta$ controls the dispersion 
of the plasmon in this system.

We emphasize that the results presented in Figs. 
\ref{intensity}-\ref{angsig} correspond to the contribution from intraband transitions that take 
place near the three independent $\mathbf{M}$-points (in this
case the concept of valley is meaningless). 
Note that Fig. \ref{distfig}  shows the effect of the anisotropic electronic 
distribution near each $\mathbf{M}$-point and the different occupations of each $\mathbf{M}$-point.
For this electronic distribution, the parity symmetry is broken (see Fig. \ref{distfig}), and, as a consequence, we can have
a finite off-diagonal conductivity. The same symmetry is broken in the in
the Hamiltonian studied by Kumar {\it et al.}\cite{Anshuman2016}. However, in this case
the parity symmetry is broken
by a circular polarized pumping field that populates each valley differently (in graphene each valley is connected by the parity symmetry).

\begin{figure}  
   \vspace{0.5cm}
   \centering
   \includegraphics[scale=0.22]{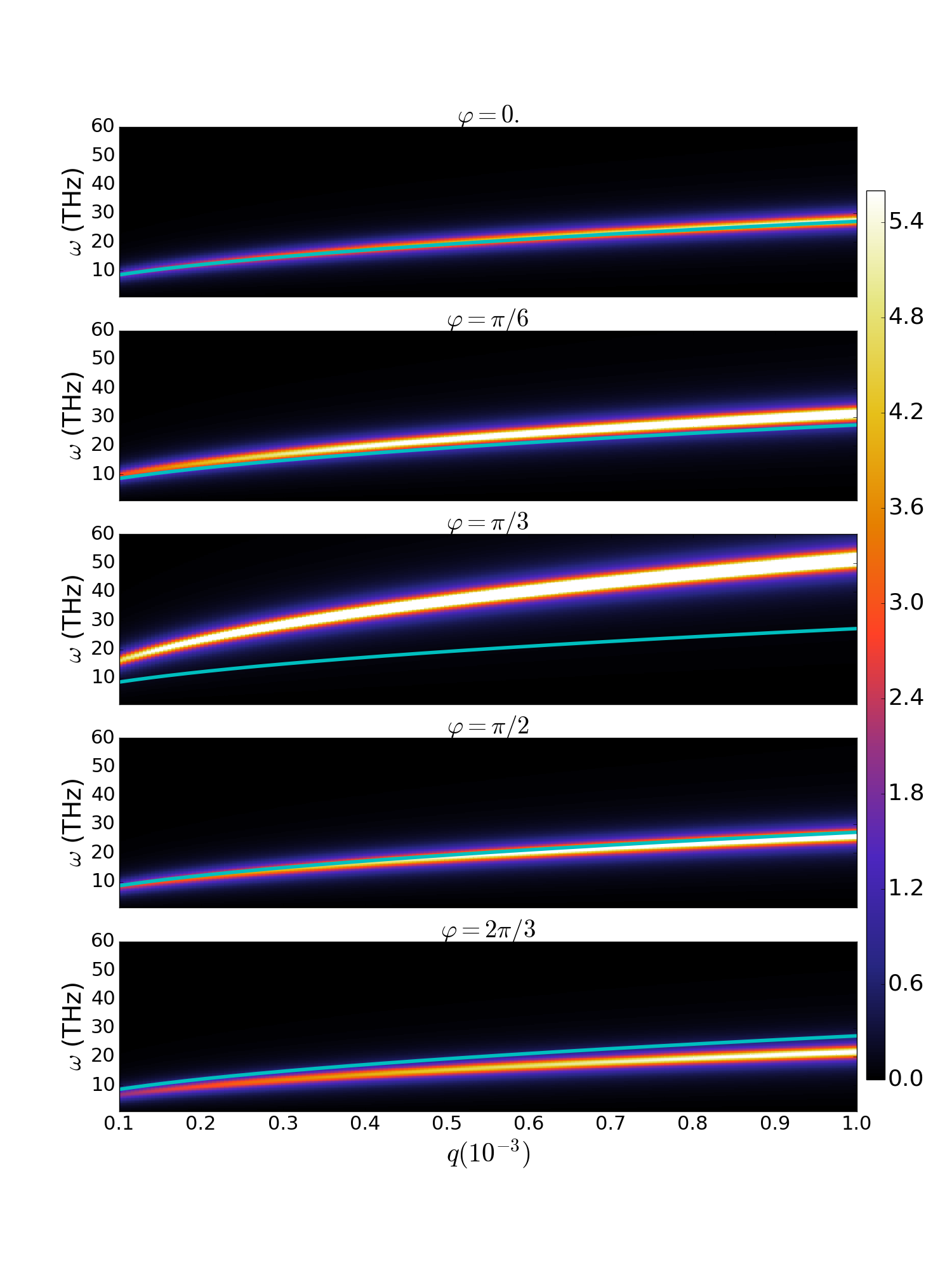}
   \caption{(Color on-line.) Loss function for different polarizations of the 
   probe field as function of the dimensionless wavenumber (multiplied by $10^3$). 
   The parameters are $\mathcal{E}_0=0.707$ GV/m, $\theta=\pi/4$, $\hbar\omega_p=2t_\text{TB}$, $E_F=0.2$ eV, $\hbar\gamma_0=14$ meV, and 
   $\hbar\gamma_p=28$ meV. The solid (cyan) curve is the plasmon dispersion for the semi-analytical result in Eq. (\ref{pdisp}) after an average of the effective Fermi energy on the polarization angle $\varphi$}  \label{listdie}
\end{figure}

One experimental way of accessing the dispersion of the plasmons in a given material is to perform
a EELS experiment. This spectroscopic technique is based on the excitation of plasmons by moving 
charges. When exciting a plasmon wave, the incoming electrons lose part of their kinetic energy.
Theoretically, the loss function, which encodes the excitation of the plasmons by the moving electrons, 
is  defined in terms of  the dielectric function as:
\be
{\cal L}(\mathbf{q},\omega)= -\Im\left\{\frac{1}{\varepsilon(\mathbf{q},\omega)}\right\}\,.
\ee
This quantity is depicted  in Fig. \ref{listdie}, for different values of the probing polarization angle
$\varphi$. The dielectric function was calculated using Eq.
(\ref{dieexp}) and the pumping susceptibility is given by Eq. (\ref{eqi}). In Fig. \ref{listdie} we can see the characteristic plasmon signature in the loss function. It is clear that the plasmon spectrum depends significantly on the polarization of the probing field, or, in other terms, on the direction of the 
momentum in the Brillouin zone. The width of the plasmon spectrum
is proportional to the relaxation rate $\gamma_0$. For making apparent the anisotropy  we also depict (solid line) the dispersion of the plasmon after an average of the effective Fermi energy on the polarization angle $\varphi$; the anisotropy is obvious.

Let us now  discuss the reason why the plasmon  characterizing  pumped graphene
out-of-equilibrium
 is similar to that of  doped graphene in equilibrium, in what concerns their small
 energy  values. In the latter case, for $\hbar\omega+\hbar v_F q<E_F$ and $\omega>v_F q$, where $q$ is the wavenumber
and $\omega$ the frequency, interband process are suppressed by Pauli-blocking
 and the susceptibility is dominated by intraband processes, where losses are proportional to the relaxation rate $\gamma_0$ (for $\gamma_0=0$ the usual plasmons are
 infinitely long-lived in this momentum-frequency window). In this regime, graphene supports plasmons with small attenuation with a relation dispersion proportional to $\sqrt{q}$. On the other hand, when we consider the case of the pumped distribution, the situation is similar, because interband process, that attenuates the plasmon, only occur for frequencies $\omega$ near the pumped frequency $\omega_p$.  Since we are considering the regime
 $\omega\ll\omega_p$, the attenuation of the plasmons of the non-equilibrium electron gas is essentially 
 controlled by the value of $\gamma_0$ (the plasmons cannot decay via particle-hole processes in this regime, as it happens in the case of an equilibrium plasma). Therefore, the correspondent pumped susceptibility is similar in the sense that the imaginary part is proportional to the scattering time [see Eq. (\ref{efsu})]. Thus, we can expect for 
 plasmons in the out-of-equilibrium electron gas the same level of attenuation of the conventional plasmons in graphene. As consequence the former anisotropic plasmons are expected to be long lived as are
 their siblings in the equilibrium electron gas. 

\begin{figure} 
   \vspace{0.5cm}
   \centering
   \includegraphics[scale=0.22]{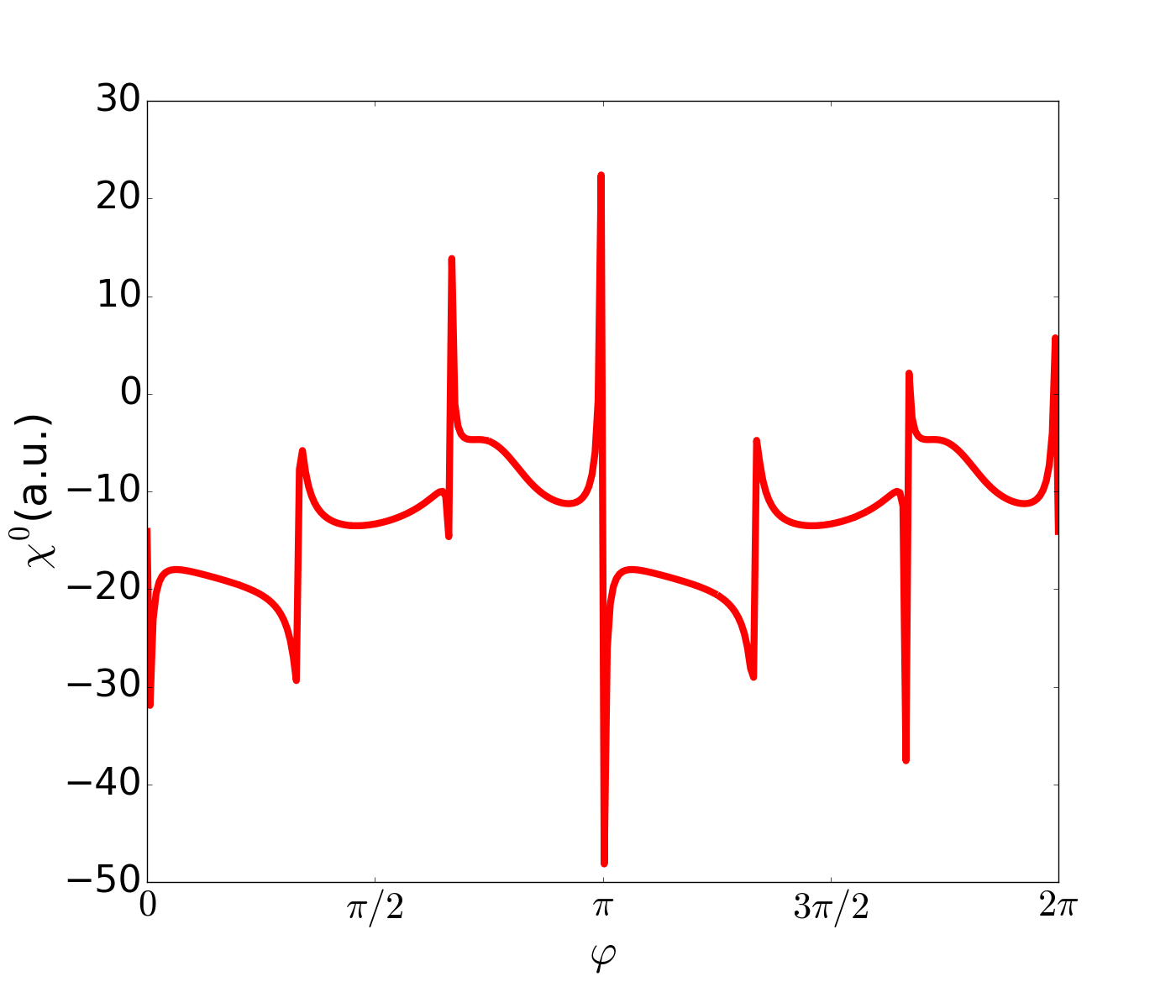}
   \caption{(Color on-line.) Static susceptibility for $q=2.10^{-3}$, $\mathcal{E}_0=0.707$ GV/m, and $\theta=\pi/4$ as function of the polarization angle of the probe field. Note the existence of points where the susceptibility is zero. Near and and at these points Mermin's approach breaksdown.}  \label{static0}
\end{figure}

In Fig. \ref{static0} we show that the graphene static susceptibility have zeros that renders the term $(\tau\omega)^{-1}\chi(q,\omega)/\chi^0(q)$ ($\tau=1/\gamma$) in Mermin's susceptibility large, even at small $q$. In this case the use of the Mermin's equation is no longer valid, since the assumption that the fluctuations of the local Fermi Energy, which are proportional to $1/\chi^0(q)$, are small is no longer true and the approximation
leading to Mermin's equation breaks down.

\begin{figure}  
   \vspace{0.5cm}
   \centering
   \includegraphics[scale=0.24]{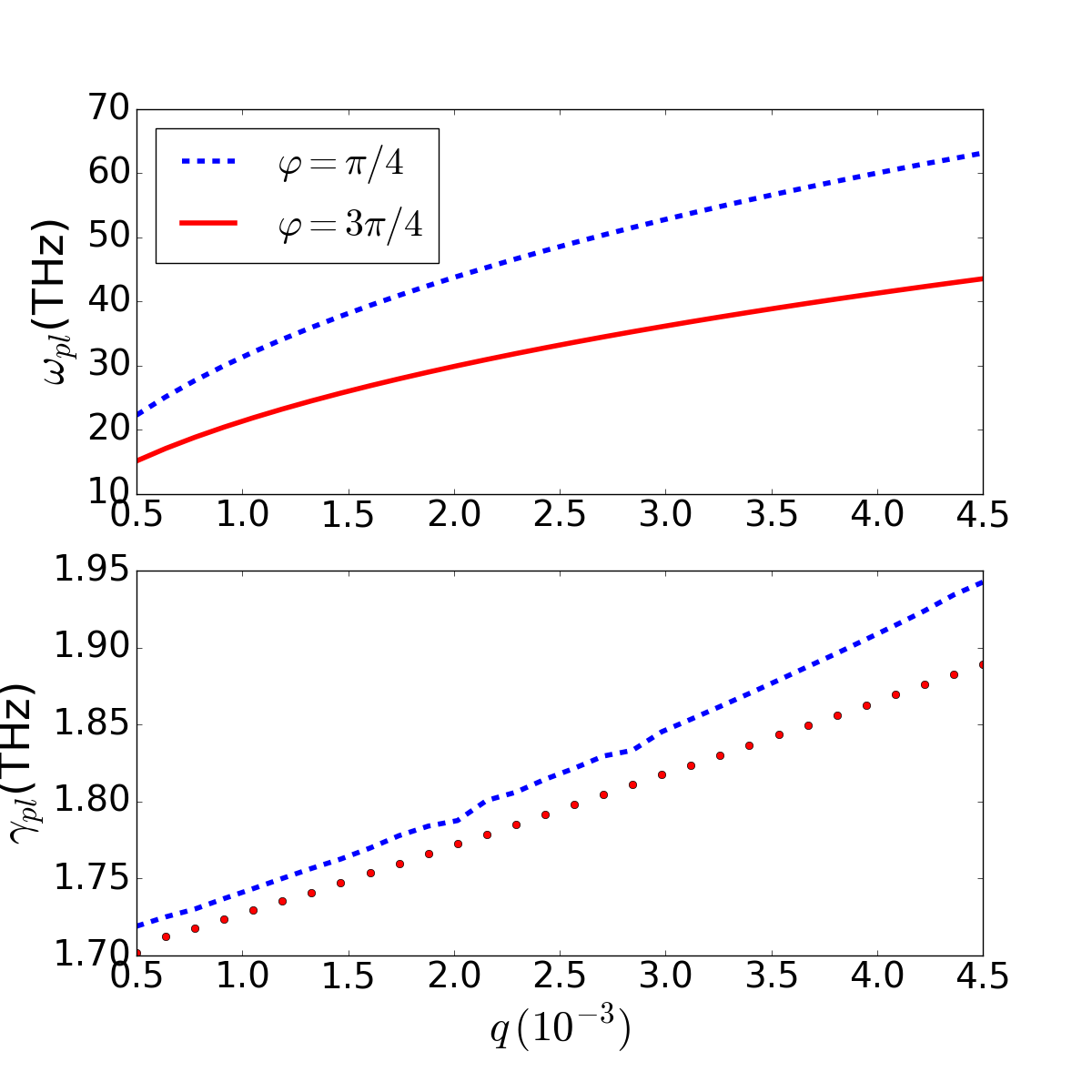}
   \caption{(Color on-line.) Plasmon dispersion relation (top panel)
   and plasmon lifetime (bottom panel) 
as function of the dimensionless wave number (multiplied by $10^3$),  
   for $\mathcal{E}_0=0.707$ GV/m, $\hbar\omega_p=2t_\text{TB}$, $\theta=\pi/4$, for two probing polarization angles. Note that $\gamma_{pl}\ll\omega_{pl}$.}  
    \label{plasmont}
\end{figure}

\section{Spectrum of the surface plasmon-polaritons in the out-of-equilibrium regime} \label{spp}

As a conductive two dimensional system, graphene supports surface plasmon-polaritons. We now want to address the propagation of the these quasi-particles on the surface of graphene due to the electron gas 
created by the pumping field.
We will see that the surface plasmon-polariton spectrum in pumped graphene shows a dispersion
strongly dependent on the $\varphi$ angle, the polarization angle of the probing field. A surface plasmon-polariton (SPP) is an hybrid particle that couples electromagnetic radiation to the 
free oscillations of an electron gas in a conductor. In graphene, the spectrum of an SPP depends critically on the nature of the optical 
conductivity of the system (for a discussion about surface plasmon-polariton in graphene see Refs. [\onlinecite{Bludov20132,Xiao2016}]).
Indeed, it can be shown that the condition for the existence of an SPP is given by\cite{Bludov20132}
\be
\left[ \frac{\varepsilon_1}{k_1}+\frac{\varepsilon_2}{k_2} +\frac{i\sigma_{xx}}{\omega \varepsilon_0}\right]
\left[\frac{ k_1+k_2 }{\omega\mu_0}-i\sigma_{yy}\right]- \frac{\sigma_{xy}\sigma_{yx}}{\omega\varepsilon_0} =0, \label{sppeq}
\ee
for a wave propagating along the $x$ direction and decaying exponentially along the direction perpendicular to the graphene plane. When $\sigma_{xy}=\sigma_{yx}=0$, the transverse electric and the transverse magnetic  modes decouple.
In the case we are considering here this is not the case, since the non-equilibrium nature of the electron gas created by the pumping 
induces a finite value for $\sigma_{xy}$.
However, since time reversal symmetry is not explicitly broken in this case, we have the condition that $\sigma_{xy} = \sigma_{yx}$.
Using the  calculated  conductivity tensor in
Eq. (\ref{sigef}) and 
the coefficients $C_{ij}$ calculated through Eq. (\ref{cdef}), the spectrum of the SPP can be obtained.

\begin{figure}  
   \vspace{0.5cm}
   \centering
   \includegraphics[scale=0.36]{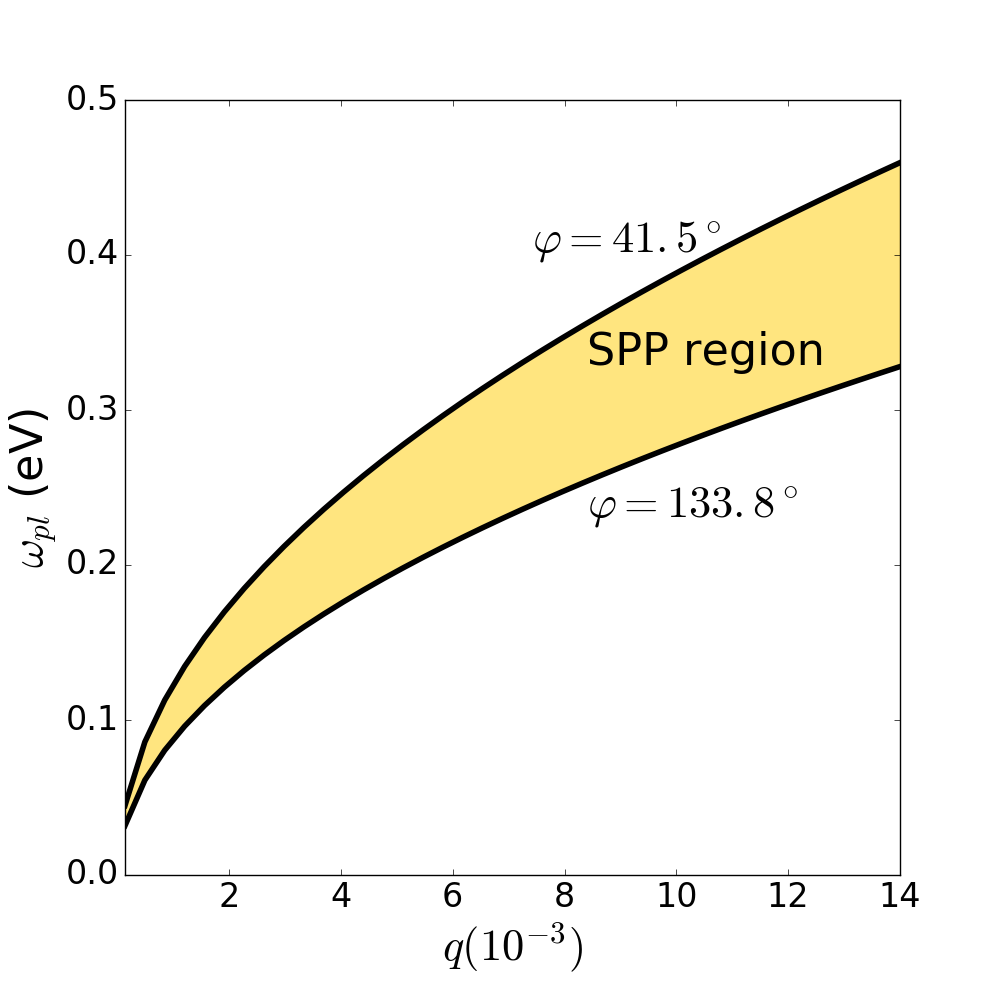}
   \caption{(Color on-line.) Surface plasmon-polaritons dispersion relation, 
   of the out-of-equilibrium electron gas,
   for different values of the angle $\varphi$ of the polarization of the probing radiation (shadded ragion)
   as function of the dimensionless wave vector (multiplied by $10^3$). Note that all the 
   angles in the interval $\varphi\in[0,2\pi]$ are contained in the shaded region. The boundary values are determined by the co-sinusoidal form of the effective Fermi energy.  Note that the variation of $\omega_{pl}$ as
   $\varphi$ varies is substantial. Therefore the anisotropy is amenable of experimental verification.
   The parameters are: $\mathcal{E}_0=0.5$ GV/m, $\hbar\omega_p=2t_\text{TB}$, $\theta=\pi/4$, $E_F=0.4$ eV, $\hbar\gamma_0=14$ meV, and $\hbar\gamma_p=28$ meV.  
   } \label{sppfig}
\end{figure}

The dispersion relation of the surface plasmon-polariton due to the non-equilibrium electron gas depends on the orientation of the direction of propagation of the wave with respect
to the crystalline lattice. To describe the propagation along another direction, we can rewrite Eq. (\ref{sppeq}) in the new reference frame or, alternatively, rotate the conductivity tensor.  
The latter can be achieved with the usual 2D rotation matrix  $M_\varphi$, $\sigma^\prime = M_\varphi \sigma M^{-1}_\varphi$:
\be
M_\varphi=\begin{pmatrix} \cos\varphi && -\sin\varphi  \\ \sin\varphi && \cos\varphi \end{pmatrix}.
\ee
In Fig. \ref{sppfig} we show, for fixed $\mathcal{E}_0$, $\omega_p$, $E_F$, and $\theta$, the surface-plasmon polariton in graphene from the solution of Eq. (\ref{sppeq}). The shaded region corresponds to different values of the variable $\varphi$, between those represented by the black solid lines at the borders of the shaded region. We  see again the strong dependence of the optical properties upon the probe angle $\varphi$, which is measured by the anisotropy in the SPP spectrum. Note that the variation
of the spectrum with $\varphi$ is quite substantial and therefore amenable to experimental verification.

\section{Final comments} \label{fc}

In this work we have considered a pump-probe problem, where the 
pumping field is a relatively intense 
and pulsed wave field, with a pulse duration much larger than 1 ps. In this situation we can reach a stationary state where an out-of-equilibrium electron gas is maintained in the conduction band in graphene. We have considered the case where the frequency of the pumping field lies in the UV-range of the electromagnetic spectrum. In this case the electrons are pumped to the $\mathbf{M}-$point in the 
Brillouin zone. In addition to the pumping, a probe field of much smaller frequency probes the out-of-equilibrium electron gas. This allows us to access the collective plasma wave ---plasmons--- in the out-of-equilibrium electron gas. We have shown that for pumping field of this frequency the excitation of the three $\mathbf{M}-$points in the Brillouin zone is uneven, at odds with the excitation of an electron gas near the 
Dirac points. This is a consequence of the strong deviation of the band structure of graphene from the Dirac cone approximation. Indeed, near the $\mathbf{M}-$point the band structure has a saddle point nature being, therefore, very different from the Dirac cone. Interestingly enough, we have found that the plasmon in the out-of-equilibrium electron gas still scales with the $\sqrt{q}$ as in the case of the Dirac plasmons. This is a consequence of the form of the charge-charge susceptibility, which scale as $q_iq_j$ ($i=x,y$) 
in the long wavelength limit 
(note that in the Dirac cone approximation the charge-charge susceptibility scales as $q^2$). This scaling can still be written in terms of $q^2$ if we introduce an effective Fermi energy, depending on the properties of the pumping field. The anisotropy of the plasmon dispersion in the Brillouin zone originates from the scaling $q_iq_j$ and is enconded in the effective Fermi energy. At the more fundamental level, the fact that the out-of-equilibrium susceptibility scales with $q_iq_j$ in the long wavelength limit 
is a consequence of the continuity equation (\ref{siseq}) that links the susceptibility with the conductivity. If in the long wavelength limit the susceptibility scales with a power lower than $q^2$ the conductivity would diverge and if the power is  greater than $q^2$, the conductivity would be null. 

Due to the relation between the charge-charge susceptibility and the optical conductivity, it is possible to define an out-of-equilibrium optical conductivity. Interestingly, the non-linear dependence of the out-of-equilibirum distribution function on the pumping field allows for a finite value of $\sigma_{xy}^\text{intra}=\sigma_{yx}^\text{intra}\ne 0$
(for the interband conductivity the situation is identical
$\sigma_{xy}^\text{inter}=\sigma_{yx}^\text{inter}\ne 0$.
Onsagar relation requires $\sigma_{xy}(H) = -\sigma_{yx}(H)$ in the presence of magnetic field $H$, or similarly broken time reversal symmetry. A Hamiltonian with circular polarized external light field is not time invariant, in which case we would have a different result from above). 
This has an impact on the 
spectrum of the surface plasmon-polaritons (SPPs) that can be supported by the out-of-equilibirum electron gas, as in this case, the TE and TM polarization are coupled to each other. We have found that the measured values for SPP spectrum depend on the orientation of the polarization of the probing field. This is a consequence of the anisotropy of the optical conductivity of graphene in the regime considered. 

What is missing form this work is a detailed study of the effect of electron-phonon and electron-electron interactions, which has been included only at the level of a phenomenological scattering rate. Therefore phenomena such as carrier multiplication is not included in our description. It would be an interesting to discuss this problem in the regime we have considered, a problem that was not analysed in the literature before, but this is outside the scope of this paper. 

Let us comment briefly on the nature of the light source needed to deliver the required electric field
intensity to observe the anisotropy. As noted in the introduction, due to excitonic effects\cite{Geim_excitonic} the position of the absorption maximum at the 
$\mathbf{M}-$point (due to inter-band transitions) is shifted from 5.4 eV to about $\sim$4.6 eV ($\lambda\sim 270$ nm), a wavelength for which there are available lasers. 
The required wavelength can be obtained from the fourth harmonic (cascade two harmonic generation) of a Q-switch diode-pump 
solid-state laser. This is a very popular wavelength,  266 nm,  which corresponds to a transition of $\sim $4.7eV, almost the excitonic resonance.
Assuming a peak power for the laser of about $P\sim65$ kW,
and a FWHM for the minimum beam waist of $w=600$ $\mu$m 
(these are figures of commercially available lasers)
it follows that the intensity of the electric field is
about
\begin{equation}
{\cal E}_0 \sim\sqrt{\frac{4P\log 2 }{\pi w^2c\epsilon_0}}\sim 0.008\, \mathrm{GV/m}\,,
\label{eq_estimation}
\end{equation} 
which is not yet enough for observing the effects we have discussed above for weakly doped graphene, which become apparent
for ${\cal E}_0\ge0.1$ GV/m. However
a lens can be used to increase the value of ${\cal E}_0$ (see ahead).
We need to comment here on the importance of the magnitude of $\gamma_0$
and $\gamma_p$ on the value of ${\cal E}_0$ needed for observing the effects we have addressed in this paper. 
In this work we have assumed that $\hbar\gamma_p=28$ meV 
(larger than $\hbar\gamma_0$)
in connection with the characteristic  time $\tau_\text{em}=0.1$ ps
implied by luminescence experiments;
if we had chosen the value of $\tau_\text{em}=0.01$ ps, corresponding
to the lowest figure suggested by experiments,\cite{Heinz2010}
 the value of ${\cal E}_0$ necessary for observing the anisotropy 
would have decreased by about $\sqrt{10}$.
We have also verified that in the opposite regime
$\gamma_0\gg\gamma_p$ the value of ${\cal E}_0$ increases significantly
over the value given by Eq. (\ref{eq_estimation}) and the observation of the 
anisotropy reported in this paper may be out of experimental reach.

Let us note here that the anisotropy is not a property of the $\mathbf{M}-$point alone, so we can choose
a larger wavelength than that necessary to excite electrons at the
$\mathbf{M}-$point 
 but still small enough for the Dirac-cone approximation to hold. In this condition, similar effects
to those described in this paper for the $\mathbf{M}-$point would still be visible. Note that our approach takes the full band structure into account and therefore is valid for all energies as long as $\omega_p\gg\omega$; the choice for the $\mathbf{M}-$point 
was only a matter of selecting a high symmetry point in the Brillouin zone. 

Another  important experimental
constrain is the threshold  power per unit area above which graphene is damaged.
It was found \cite{destroy} that for a laser of $\lambda=248$ nm the threshold power per unit area was of the order of $3.8\times10^{10}$ W/m$^2$
(The experiment considered 500 laser pulses of 20 ns duration and a repetition rate of 100 Hz; see however
Ref. [\onlinecite{destroy2}]). This value is an order of magnitude smaller than
$P/w^2\sim1.8\times10^{11}$ W/m$^2$ estimated from the numbers
given above. This implies a reduction in the power $P$ and therefore
a smaller value of ${\cal E}_0$. Indeed, using $P/w^2\sim3.8\times10^{10}$ W/m$^2$ it follows
\begin{equation}
{\cal E}_0\sim 0.004\,\mathrm{GV/m}\,,
\end{equation}
which is only a factor of 2 smaller than that found in Eq. (\ref{eq_estimation}). 
On the other hand, a single pulse of 20 ns duration is much larger than  1 ps needed for attaining the steady state (see Fig. \ref{rtime}). Therefore, within the duration of a
single 20 ns pulse it is possible to excite the plasmon in the out-of-equilibrium gas and measure their existence. The use of a single
 pulse allows to multiply the power value of  $3.8\times10^{10}$ W/m$^2$ by 500
(number of repetitions needed to observe clear changes in the
D-peak of the Raman spectrum of graphene) thus allowing  a value of ${\cal E}_0$, for a single pulse, of the order of 
\begin{equation}
{\cal E}_0\sim 0.1\,\mathrm{GV/m}\,,
\end{equation}
and introducing only a small amount of laser ablation on graphene.
Therefore, the value used in the paper of ${\cal E}_0\sim 0.7\,\mathrm{GV/m}$ is acceptable for illustrative purposes, since
it is about 7 times higher 
than the estimation made above. Therefore, the used value for
${\cal E}_0$  has been chosen for making the effects more apparent to the naked eye. Naturally, focusing 
the laser spot onto an area smaller than $w^2$,
makes the figure of $0.7\,\mathrm{GV/m}$ acceptable.
Indeed, for the laser indicated above with an exit spot  waist of
$w=800$ $\mu$m we can focus it down to a spot waist of 
$12$ $\mu$m which implies that the value of ${\cal E}_0\sim 0.7\,\mathrm{GV/m}$ in well within the experimental range. In conclusion,
it is expected that by lowering the repetition rate, the damage threshold would increase beyond the $3.8\times10^{10}$ W/m$^2$ reported in Ref.
[\onlinecite{destroy}]. This is justified by the fact that the average power incident on the sample will decreased and thus, also the damage due to thermal effects. Furthermore, focusing the laser beam leads to values of 
${\cal E}_0$  more than one order of magnitude larger that given by Eq. (\ref{eq_estimation}), implying that
the effects predicted in this paper become experimentally accessible.

Let us finally 
comment on a
difference between our results and those of Ref. [\onlinecite{Agarwal2016}]:
the Rabi frequency  in Eq. (\ref{rabi}) depends on the gauge choice for the electromagnetic field. When we add the relaxation time
we break gauge invariance. In the work of Singh {\it et al.} \cite{Agarwal2016} the equations of motion were obtained with the minimal coupling in the Couloumb gauge. The corresponding Rabi frequency has a factor $\omega_\mathbf{k}/\omega$ in comparision with Eq. (\ref{rabi}). However,  when  the relaxation time goes to infinity, $\gamma\rightarrow0$, the  two approachs give identical results. Indeed, from the equations in Appendix \ref{coc}, that were written in the same gauge as that used in Ref. [\onlinecite{Agarwal2016}], we can make $\mathbf{q}_0=0$ (the wave number of the probing field) and proceed in the same way as we did in  Sec. \ref{neqdm} to obtain the same results as those found in Ref.
[\onlinecite{Agarwal2016}].

\section*{Acknowledgments}
A. J. Chaves acknowledges the scholarship from the Brazilian
agency  CNPq (Conselho Nacional de Desenvolvimento
Cient\'ifico e Tecnol\'ogico).
 N.M.R. Peres acknowledges support from the European Commission through
the project ``Graphene-Driven Revolutions in ICT and Beyond"
(Ref. No. 696656) and the Portuguese Foundation for Science and Technology 
(FCT) in the framework of the Strategic Financing UID/FIS/04650/2013.
The authors acknowledge José Carlos Viana Gomes for discussions
that led to the estimations made in Sec. \ref{fc}.

\begin{appendix} 
\section{Tight-binding model for graphene subjected to an external electric field}
\label{app_tb}
\begin{figure}  
   \vspace{0.5cm}
   \centering
   \includegraphics[width=8cm]{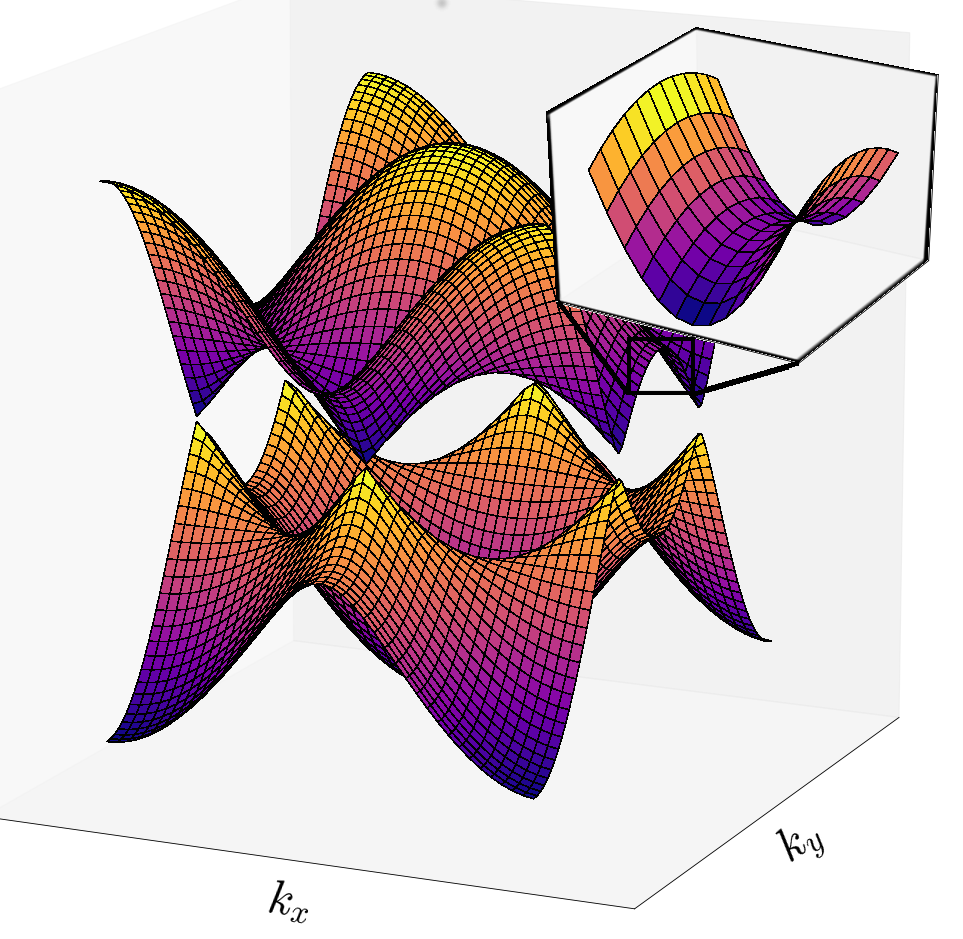}
   \caption{(Color on-line.)      
Electronic band structure of graphene for $\pi-$electrons. The zoom-in shows the band structure around the $\mathbf{M}-$point in the Brillouin zone.
The saddle-like nature of the band structure is clearly visible around that point.    
   } \label{fig_full_spectrum}
\end{figure}

The graphene tight-binding Hamiltonian in second quantization reads:
\be
 H_0= \sum_{i,n} t_\text{TB} \,\hat{a}^\dagger_{\mathbf{R}_n}\hat{b}_{\mathbf{R}_n+\boldsymbol{\delta}_i}+\text{h.c.}, \label{H0}
\ee
where $\hat{a}^\dagger_{\mathbf{R}_n}$ and $\hat{b}_{\mathbf{R}_n+\boldsymbol{\delta}_i}$ obey anti-commutation relations
and $\boldsymbol{\delta}_i$ are the nearest neighbors vectors connecting
an atom in sub-lattice $A$ to another one in sub-lattice $B$ [see Eq. (\ref{eq_deltas})].
We can define the Fourier transform and its inverse as:
\begin{subequations}
\bea
\hat{a}_\mathbf{k}&=&\frac{1}{\sqrt{N_c}}\sum_n e^{-i\mathbf{k} \cdot\mathbf{R}_n} \hat{a}_{\mathbf{R}_n},\\
\hat{b}_\mathbf{k}&=&\frac{1}{\sqrt{N_c}}\sum_n e^{-i\mathbf{k} \cdot\left(\mathbf{R}_n+\boldsymbol{\delta}_i\right)} \hat{b}_{\mathbf{R}_n+\boldsymbol{\delta}_i},
\eea
\end{subequations}
\begin{subequations}
\bea
\hat{a}_{\mathbf{R}_n}&=&\frac{1}{\sqrt{N_c}}\sum_{\mathbf{k}\in 1º B.Z.} e^{i\mathbf{k} \cdot\mathbf{R}_n}\hat{a}_\mathbf{k} ,
\\
\hat{b}_{\mathbf{R}_n+\boldsymbol{\delta}_i}& =&\frac{1}{\sqrt{N_c}}\sum_{\mathbf{k}\in 1º B.Z.} e^{i\mathbf{k} \cdot\left(\mathbf{R}_n+\boldsymbol{\delta}_i\right)} \hat{b}_\mathbf{k},
\eea \label{mombasis}
where the sum over $n$ is performed over the entire lattice and the sum in $\mathbf{k}$ is performed over the first Brillouin zone.

\end{subequations}
After a Bogoliubov transformation the basis that diagonalize $H_0$ is:
\begin{subequations}
\bea
\hat{c}_{\mathbf{k}}= \frac{e^{i\varphi_\mathbf{k}}}{\sqrt{2}}\left(   \hat{a}_\mathbf{k}+e^{i\Theta_{\mathbf{k}}}\hat{b}_\mathbf{k}\right),
\\
\hat{d}_{\mathbf{k}}= \frac{e^{i\varphi_\mathbf{k}}}{\sqrt{2}}\left(   \hat{a}_\mathbf{k}-e^{i\Theta_{\mathbf{k}}}\hat{b}_\mathbf{k}\right),
\eea \label{bogtr}
\end{subequations}
and the inverse transformation reads:
\begin{subequations}
\bea
\hat{a}_\mathbf{k}&=&\frac{e^{-i\varphi_\mathbf{k}} }{\sqrt{2}} \left(\hat{c}_\mathbf{k}+\hat{d}_\mathbf{k}\right),
\\
\hat{b}_\mathbf{k}&=&\frac{e^{-i\varphi_\mathbf{k}}e^{-i\Theta_\mathbf{k}} }{\sqrt{2}} \left(\hat{c}_\mathbf{k}-\hat{d}_\mathbf{k}\right),
\eea
\end{subequations}
where $\varphi_\mathbf{k}$ is a global arbitrary phase. The phase $\Theta_\mathbf{k}$ is the argument of
\be
\phi_\mathbf{k}=\sum_{i=1}^3 e^{i\mathbf{k} \cdot \boldsymbol{\delta}_i},
\ee
that is, $\Theta_\mathbf{k}=\text{arg }\phi_\mathbf{k}$. In the basis (\ref{bogtr}) the Hamiltonian $H_0$ is written as:
\be
H_0 = \sum_\mathbf{k} E_\mathbf{k} \left( \hat{c}^\dagger_\mathbf{k} \hat{c}_\mathbf{k}- \hat{d}^\dagger_\mathbf{k} \hat{d}_\mathbf{k} \right)\,.
\ee
The band structure given by $\pm E_\mathbf{k}$ is depicted in Fig. \ref{fig_full_spectrum} together with a zoom-in around the $\mathbf{M}-$point.
For writing the  the interaction term with the electric field we need the position operator written as:
\begin{subequations}
\bea
\mathbf{\hat{R}}_A&=&\sum_n \mathbf{R}_n \hat{a}^\dagger_{\mathbf{R}_n} \hat{a}_{\mathbf{R}_n},
\\
\mathbf{\hat{R}}_B&=&\sum_n \left(\mathbf{R}_n+\boldsymbol{\delta_1}\right) \hat{b}^\dagger_{\mathbf{R}_n+\boldsymbol{\delta}_1} \hat{b}_{\mathbf{R}_n+\boldsymbol{\delta}_1},
\eea \label{positop}
\end{subequations}
which in the basis given by Eq. (\ref{bogtr}) it reads:
\bea
\mathbf{R}=-\frac{(2\pi)^2i}{2N_c a_0^2} \sum_{\mathbf{k},\mathbf{q}}\left[\boldsymbol{\nabla}_\mathbf{q}\delta\left(\mathbf{q}\right)\right]
e^{i(\varphi_\mathbf{k+q} -\varphi_\mathbf{k} )} \nonumber \\ \left[\left( 1 +e^{i( \Theta_{\mathbf{k+q}}-\Theta_\mathbf{k})} \right)\left( \hat c^\dagger_\mathbf{k+q}\hat c_\mathbf{k} +\hat d^\dagger_\mathbf{k+q}\hat d_\mathbf{k}\right)+\right. \nonumber\\ \left. +\left(   1-e^{i( \Theta_{\mathbf{k+q}}-\Theta_\mathbf{k})}\right) \left(\hat c^\dagger_\mathbf{k+q}\hat d_\mathbf{k}+\hat d^\dagger_\mathbf{k+q}\hat c_\mathbf{k}\right) \right]\,,
\eea
where $\delta\left(\mathbf{q}\right)$ is the Dirac delta-function of zero momentum.
Therefore, the radiation-electron interaction is finally written as:
\begin{flalign}
H_I= e\boldsymbol{\mathbf{E}}\cdot \sum_\mathbf{k}
\left[i\nabla_\mathbf{k}\left(\hat{n}_{c,\mathbf{k}}+\hat{n}_{v,\mathbf{k}}\right)+\frac{\nabla_\mathbf{k} \Theta_\mathbf{k}}{2}\left(\hat{p}_{cv,\mathbf{k}}+\hat{p}_{vc,\mathbf{k}}\right)\right],
\end{flalign}
where we have defined:
\begin{subequations}
\bea
\hat{n}_{c,\mathbf{k}}&=&c^\dagger_\mathbf{k}c_\mathbf{k},
\\
\hat{n}_{v,\mathbf{k}}&=&d^\dagger_\mathbf{k}d_\mathbf{k},
\\
\hat{p}_{cv,\mathbf{k}}&=& c^\dagger_\mathbf{k}d_\mathbf{k},
\\
\hat{p}_{vc,\mathbf{k}}&=&d^\dagger_\mathbf{k}c_\mathbf{k},
\\
\varphi_{\mathbf{k+q}}&=&-\frac{\Theta_{\mathbf{k+q}}} {2}.
\eea
\end{subequations}

\section{Derivation of the Bloch equations for graphene} \label{aprtime}

To obtain the Bloch equations in graphene we calculate the commutator (\ref{heq}) with the introduction of two 
phenomenological damping terms and with the density matrix written in the basis (\ref{bogtr}). After the calculation of the
expectation values we obtain the set of equations:
\begin{subequations}
\bea
-\partial_t n_{c,\mathbf{k}}=\gamma_0\left(  n_{c,\mathbf{k}}-f_{c,\mathbf{k}}\right)+i\Omega_\mathbf{k}(t)\Delta p_\mathbf{k}, \label{n1}
\\
-\partial_t n_{v,\mathbf{k}}=\gamma_0\left(  n_{v,\mathbf{k}}-f_{v,\mathbf{k}}\right)-i\Omega_\mathbf{k}(t)\Delta p_\mathbf{k}, \label{n2}
\\
\left(\partial_t+i\omega_\mathbf{k}+\gamma_p\right) p_{cv,\mathbf{k}}=-i\Omega_\mathbf{k}(t)\Delta n_\mathbf{k}, \label{p1}
\\
\left(\partial_t-i\omega_\mathbf{k}+\gamma_p\right) p_{vc,\mathbf{k}}=i\Omega_\mathbf{k}(t) \Delta n_\mathbf{k}, \label{p2}
\eea\label{alleq2}\end{subequations}
where $\hbar\omega_\mathbf{k}=2E_\mathbf{k}$, $\Delta n_\mathbf{k}=n_{c,\mathbf{k}}-n_{v,\mathbf{k}}$, $\Delta p_\mathbf{k}=p_{cv,\mathbf{k}}-p_{vc,\mathbf{k}}$, $f_{c/v,\mathbf{k}}$ is the Fermi-Distribution for the conduction/valence band, and $\gamma_0$($\gamma_p$) is a relaxation term. The time dependence on $n_{c/v,\mathbf{k}}$, $p_{vc/cv,\mathbf{k}}$, and $\boldsymbol{\mathcal{E}}$ is omitted, and we have defined:
\be
\Omega_\mathbf{k}(t)=\frac{e a_0\boldsymbol{\mathcal{E}}(t)\cdot \boldsymbol{\nabla}_{\mathbf{k}} \Theta_\mathbf{k}}{2\hbar}\,.
\ee
Summing Eqs. (\ref{n1}) and (\ref{n2}) we obtain:
\be
\partial_t (n_{c,\mathbf{k}}+n_{v,\mathbf{k}}) = - \gamma_0\left(n_{c,\mathbf{k}}+n_{v,\mathbf{k}}-(f_{c,\mathbf{k}}+f_{v,\mathbf{k}}) \right),
\ee
which has the exact solution:
\be
n_{c,\mathbf{k}}(t)+n_{v,\mathbf{k}}(t) = c(\mathbf{k}) e^{-\gamma_0t}+ f_{c,\mathbf{k}}+f_{v,\mathbf{k}}, 
\ee
where $c(\mathbf{k})$ depends on the initial conditions. 

For a system that is initially in thermal equilibrium, $c(\mathbf{k})=0$ and:
\be
n_{c,\mathbf{k}}(t)+n_{v,\mathbf{k}}(t)= f_{c,\mathbf{k}}+f_{v,\mathbf{k}}, \label{ax1e}
\ee
thus we introduce the deviation $\rho_\mathbf{k}(t)$ through:
\begin{subequations}
\bea
n_{c,\mathbf{k}}(t)=f_{c,\mathbf{k}}+\rho_\mathbf{k}(t),
\\
n_{v,\mathbf{k}}(t)=f_{v,\mathbf{k}}-\rho_\mathbf{k}(t)\,.
\eea \label{ax2e}
\end{subequations}
We also  note that the complex conjugate of (\ref{p2}) reads:
\be
\left(\frac{\partial}{\partial t}+i\omega_\mathbf{k}+\gamma_p\right) p^*_{vc,\mathbf{k}}=-i\frac{e\boldsymbol{\mathcal{E}} \cdot\nabla_\mathbf{k} \Theta_\mathbf{k}}{2}   \Delta n_\mathbf{k}, \label{p3}
\ee
and using Eq. (\ref{p1})
\be
\left(\frac{\partial}{\partial t}+i\omega_\mathbf{k}+\gamma_p\right) \left( p_{cv,\mathbf{k}}(t)-p^*_{vc,\mathbf{k}}(t)\right)=0, \label{p4}
\ee
we find the solution:
\be
p_{cv,\mathbf{k}}(t)=p^*_{vc,\mathbf{k}}(t)+c_1(\mathbf{k}) e^{(-i\omega_\mathbf{k}-\gamma_p)t},
\ee
where again $c_1(\mathbf{k})$ depends on the initial conditions.
If we assume  that the system is initially in thermal equilibrium it follows that $c_1(\mathbf{k})=0$ and:
\be
p_{vc,\mathbf{k}}(t)=p^*_{cv,\mathbf{k}}(t)=x_\mathbf{k}(t)+iy_\mathbf{k}(t)\,.
\ee

The set of four complex equations (\ref{alleq}) can be reduced to a set of three real equations for the functions $x_\mathbf{k}$, $y_\mathbf{k}$, and $\rho_\mathbf{k}$.
From Eqs. \ref{alleq} we have:
\begin{subequations}
\be
\partial_t \rho_\mathbf{k}=-\gamma_0 \rho_\mathbf{k}+4 v_\mathbf{k}(t) y_\mathbf{k},
\ee
\be
\left(\partial_t-i\omega_\mathbf{k}+\gamma_p\right) \left[x_\mathbf{k}+iy_\mathbf{k}\right]=i \Omega_\mathbf{k}(t)   \left(\Delta n^0_\mathbf{k}+2\rho_\mathbf{k}\right), 
\ee
\end{subequations}
from which finally follows that:
\begin{subequations}
\bea
 \dot{x}_\mathbf{k}&=&-\gamma_p x_\mathbf{k} -\omega_\mathbf{k}y_\mathbf{k},
\\
 \dot{y}_\mathbf{k}&=&\omega_\mathbf{k}x_\mathbf{k}-y_\mathbf{k}\gamma_p-\Omega_\mathbf{k}(t)\left(2\rho_\mathbf{k}+ \Delta n^0_\mathbf{k}\right),
\\
\dot{\rho}_\mathbf{k}&=&-\gamma_0 \rho_\mathbf{k} +2\Omega_\mathbf{k}(t) y_\mathbf{k}.
\eea
\end{subequations}
These latter set of equations is the one we have solved in the bulk of the paper.

\section{Steady-state equations for the distribution functions under continuous pumping} \label{apdem}

With the assumption that $\partial_t n^c_\mathbf{k}=\partial_t n^v_\mathbf{k}=0$ and considering a monochromatic incident field
with frequency $\omega_p$, we can write Eqs. (\ref{alleq}) as:
\begin{subequations}
\bea
\gamma_0\left(  n_{c,\mathbf{k}}-n^0_{c,\mathbf{k}}\right)+i \left\langle \Omega_\mathbf{k}(t)\Delta p_\mathbf{k} \right\rangle_t=0, \label{c1}
\\
\gamma_0\left(  n_{v,\mathbf{k}}-n^0_{v,\mathbf{k}}\right)-i \left \langle \Omega_\mathbf{k}(t)\Delta p_\mathbf{k} \right\rangle_t=0, \label{c2}
\\
\left(\partial_t+i\omega_\mathbf{k}+\gamma_p\right) p_{cv,\mathbf{k}}=-i\Omega_\mathbf{k}(t)\Delta n_\mathbf{k} \label{c3},
\\
\left(\partial_t-i\omega_\mathbf{k}+\gamma_p\right) p_{vc,\mathbf{k}}=i\Omega_\mathbf{k}(t) \Delta n_\mathbf{k}, \label{c4}
\eea\end{subequations}
where we use $\langle \rangle_t$ for time average. The solution to this set of equations is of the form:
\begin{subequations}
\bea
p_{cv,\mathbf{k}},(t)=A_1(\omega_p)e^{i\omega_p t}+B_1(\omega_p) e^{-i\omega_p t},
\\
p_{vc,\mathbf{k}}(t)=A_2(\omega_p)e^{i\omega_p t}+B_2(\omega_p) e^{-i\omega_p t},
\eea
\end{subequations}
where $A_i$, $B_i$ can be obtained from Eqs. (\ref{c3}) and (\ref{c4}) as:
\begin{subequations}
\bea
A_1(\omega_p)= \frac{  n_{c,\mathbf{k}}-n_{v,\mathbf{k}} }{\omega_p+\omega_\mathbf{k}-i\gamma_p}\frac{\bar{\Omega}_\mathbf{k}}{2},
\\
B_1(\omega_p)= \frac{  n_{c,\mathbf{k}}-n_{v,\mathbf{k}} }{-\omega_p+\omega_\mathbf{k}-i\gamma_p}\frac{\bar{\Omega}^*_\mathbf{k}}{2},
\\
A_2(\omega_p)= \frac{  n_{v,\mathbf{k}}-n_{c,\mathbf{k}} }{\omega_p-\omega_\mathbf{k}-i\gamma_p}\frac{\bar{\Omega}_\mathbf{k}}{2},
\\
B_2(\omega_p)= \frac{  n_{v,\mathbf{k}}-n_{c,\mathbf{k}} }{-\omega_p-\omega_\mathbf{k}-i\gamma_p}\frac{\bar{\Omega}^*_\mathbf{k}}{2},
\eea
\end{subequations}
with:
\be
\bar{\Omega}_\mathbf{k}=\frac{e a_0\boldsymbol{\mathcal E}_0\cdot \boldsymbol{\nabla}_\mathbf{k} \Theta_\mathbf{k}}{2\hbar}\,. \label{rabi_indep}
\ee
If we define:
\be
\alpha_\mathbf{k}=  \tau_0 \tau_p|\bar{\Omega}_\mathbf{k}|^2\frac{1+\tau_p^2\left(\omega_\mathbf{k}^2+\omega_p^2 \right)}{\tau_p^4(\omega^2_p-\omega^2_\mathbf{k})^2+2\tau_p^2(\omega^2_p+\omega_\mathbf{k}^2)+1}, \label{alphak}
\ee
with $\tau_0=1/\gamma_0$, $\tau_p=1/\gamma_p$,  we have from Eqs. (\ref{c1}) and (\ref{c2}) that:
\begin{subequations}
\be
n_{c,\mathbf{k}}-f_{c,\mathbf{k}}= \alpha_\mathbf{k} (n_{v,\mathbf{k}}-n_{c,\mathbf{k}}),
\ee
\be
n_{v,\mathbf{k}}-f_{v,\mathbf{k}}= \alpha_\mathbf{k} (n_{c,\mathbf{k}}-n_{v,\mathbf{k}}).
\ee \label{steady_state_sol}
\end{subequations}
Note that expression for $\alpha_\mathbf{k}$ is well defined even taking the collisionless regime
$(\tau_0,\tau_p)\rightarrow\infty$.

\section{Expressions for the gradient of the phase $\Theta_\mathbf{k}$}

In this appendix we present some useful functions that appear in the rest of the paper. First we 
recall the function defined in Eq. (\ref{phifunc}):
\be
\phi_\mathbf{k}=\sum_{j=1}^3 e^{i\mathbf{k}\cdot\boldsymbol{\delta}_j}, \label{phiaps}
\ee
where we have used the following choice of vectors for the orientation 
of the nearest neighbor hopping:
\begin{subequations}
\bea
\boldsymbol{\delta}_1&=&a_0 (1,0), 
\\
\boldsymbol{\delta}_2&=&a_0/2 \left(-1,-\sqrt{3}\right), 
\\
\boldsymbol{\delta}_3&=&a_0/2 \left(-1,\sqrt{3}\right).
\eea
With this choice of vectors, the eigenvalues of $H_0$ are the solution of:
\be
\left( \frac{E_\mathbf{k}}{t_{TB}}\right)^2=1 +4 \cos\left(\frac{3}{2} k_x\right) \cos\left(\frac{\sqrt{3}}{2}k_y\right)+4\cos^2\left( \frac{\sqrt{3}}{2}k_y\right),
\ee
\label{eq_deltas}
\end{subequations}
and the $\Theta_\mathbf{k}$ function  (the argument of $\phi_\mathbf{k}$)  is written as:
\be
\tan \Theta_\mathbf{k}= \frac{\sin k_x-2\sin\frac{k_x}{2} \cos\frac{\sqrt{3}k_y}{2}}{ \cos k_x+2\cos\frac{k_x}{2} \cos\frac{\sqrt{3}k_y}{2}}\,. \label{tth}
\ee
We can calculate $\boldsymbol{\nabla}_\mathbf{k}\Theta_\mathbf{k}$ through:
\be
\boldsymbol{\nabla}_\mathbf{k}\Theta_\mathbf{k}= \frac{u\boldsymbol{\nabla}_\mathbf{k} v-v\boldsymbol{\nabla}_\mathbf{k} u}{E(\mathbf{k})^2},
\ee
where we have split the function $\phi_\mathbf{k}$ in Eq. (\ref{phiaps}) into real and imaginary parts $\phi_\mathbf{k}=u+iv$. It then follows that we can obtain
the components of the gradient of the $\Theta_\mathbf{k}$ function as:
\begin{subequations}
\be
\partial_{k_x} \Theta_\mathbf{k}=\frac{1-2\cos^2\left(\frac{\sqrt{3}}{2} k_y\right)+\cos\left(\frac{3}{2}k_x\right)\cos\left(\frac{\sqrt{3}}{2}k_y\right)}{1+4 \cos\left(\frac{3}{2} k_x\right) \cos\left(\frac{\sqrt{3}}{2}k_y\right)+4\cos^2\left( \frac{\sqrt{3}}{2}k_y\right)},
\ee
\be
\partial_{k_y} \Theta_\mathbf{k}=\frac{\sqrt{3}\sin\left(\frac{3}{2}k_x\right)\sin\left( \frac{\sqrt{3}}{2}k_y\right)}{1+4 \cos\left(\frac{3}{2} k_x\right) \cos\left(\frac{\sqrt{3}}{2}k_y\right)+4\cos^2\left( \frac{\sqrt{3}}{2}k_y\right)}.
\ee
\end{subequations}
\section{Semi-analytical formula for the charge-charge correlation function} \label{saf}

In the long wavelength limit, the susceptibility for finite frequency is written in power 
of $q^2$. If we expand $\rho_{\mathbf{k}+\mathbf{q}}$ 
and $\omega_{\mathbf{k},\mathbf{q}}$ until order $q^2$, we have:
\be
\chi_\text{pump}^\text{intra}(\mathbf{q},\omega)=\frac{2e}{\hbar a_0^2}\sum_\lambda \int \frac{d^2\mathbf{k}}{(2\pi)^2} \frac{\lambda\boldsymbol{\nabla}_\mathbf{k}\rho_\mathbf{k}\cdot\mathbf{q}}{\omega-\lambda\boldsymbol{\nabla}_\mathbf{k} E_\mathbf{k}\cdot\mathbf{q}+i\gamma_0},
\ee
where we used that $N(\mathbf{k},\mathbf{q})=1+{\cal O} (q^2)$. Expanding also the denominator we find:
\be
\chi_\text{pump}^\text{intra}(\mathbf{q},\omega)=\frac{4e}{\hbar a_0^2}\frac{1}{(\omega+i\gamma_0)^2}\int \frac{d^2\mathbf{k}}{(2\pi)^2} \boldsymbol{\nabla}_\mathbf{k}\rho_\mathbf{k}\cdot\mathbf{q}\boldsymbol{\nabla}_\mathbf{k} E_\mathbf{k} \cdot \mathbf{q} , \label{ap1c}
\ee
and thus we can define:
\be
C_{ij}= \int \frac{d^2\mathbf{k}}{(2\pi)^2}\partial_i\rho_\mathbf{k}\partial_j E_\mathbf{k}\,.\label{cdef}
\ee
In terms of $C_{ij}$ we rewrite Eq. (\ref{ap1c}) as:
\be
\chi_\text{pump}^\text{intra}(\mathbf{q},\omega)= \frac{4e}{\hbar a_0^2} \frac{1}{(\omega+i\gamma_0)^2}\sum_{ij} C_{ij} q_i q_j\,.\label{chip0}
\ee
Making $q_x=q \cos\varphi$ and $q_y= q\sin\varphi$ it follows that:
\begin{align}
\sum_{ij} C_{ij} q_i q_j&= C_{xx}\cos^2\varphi+C_{yy}\sin^2\varphi
\nonumber\\
&+(C_{xy}
+C_{yx})\sin\varphi\cos\varphi\,. \label{chip}
\end{align}
With an integration by parts we can show from Eq. (\ref{cdef}) that $C_{xy}=C_{yx}$. Using trigonometric identities we can
write Eq. (\ref{chip}) as:
\be
\sum_{ij} C_{ij} q_i q_j= \frac{C_{xx}+C_{yy}}{2}+\frac{C_{xx}-C_{yy}}{2} \cos2\varphi+C_{xy}\sin2\varphi\,.\label{chip2}
\ee
Defining:
\begin{subequations}
\be
f_0=\frac{1}{\pi}\frac{C_{xx}+C_{yy}}{2},
\ee
\be
f_m=\frac{1}{\pi}\sqrt{\left(\frac{C_{xx}-C_{yy}}{2}\right)^2+C_{xy}^2},
\ee
\be
\phi=\arctan \frac{2C_{xy}}{C_{xx}-C_{yy}},
\ee
\end{subequations}
the susceptibility (\ref{chip0}) is written as:
\be
\chi_\text{pump}^\text{intra}(q,\varphi,\omega)=\frac{4e\left(f_0+f_m\cos(2\varphi-\phi) \right)}{\hbar^2 a_0^2\pi}\frac{q^2}{(\omega+i\gamma_0)^2}\,.
\ee
It is then possible to defined an effective Fermi energy $E^\text{eff}_F(\varphi)=E_F+f_0+f_m\cos(2\varphi-\phi) $, which allows to write the total susceptibility  as:
\be
\chi(q,\varphi,\omega)=\frac{4eE^\text{eff}_F(\varphi)}{\hbar^2a_0^2\pi}\frac{q^2}{(\omega+i\gamma_0)^2}.
\ee
The last result has the same functional form on frequency as that of the charge-charge susceptibility in the independent
electron gas model.
\section{Current-current response function in the out-of-equilibrium regime}\label{coc}

The tight-binding Hamiltonian for a system with two atoms per unit cell (and thus two sub-lattices A and B), and
considering only nearest neighbor hoping, is given by:
\be
H= \sum_{i,n} t_{i,n}(\mathbf{A}) \hat{a}^\dagger_{\mathbf{R}_n}\hat{b}_{\mathbf{R}_n+\boldsymbol{\delta}_i}+\text{h.c.}\,.
\ee
where we have 
Introduced the  Peierls substituion as:
\be
t_{i,n}(\mathbf{A})=t_\text{TB} \exp\left(-i \frac{e \mathbf{A}(\mathbf{R}_n)\cdot\boldsymbol{\delta}_i}{\hbar}\right),
\ee
which is 
valid when the vector potential $\mathbf{A}(\mathbf{R}_n)$ changes smoothly with the position in the lattice. The vectors $\boldsymbol{\delta}_i$  connect two nearest neighbors atoms from different sub-lattices.

Expanding the Hamiltonian in the field $\mathbf{A}$ up to second order
we obtain:
\be
H=H_0+V_1+V_2, 
\label{eq_HV1V2}
\ee
where $H_0$ is the usual tight-binding hamiltonian (\ref{H0}):
\be
H_0=t_\text{TB}\sum_{i,n} \hat{a}^\dagger_{\mathbf{R}_n}\hat{b}_{\mathbf{R}_n+\boldsymbol{\delta}_i}+\text{h.c.},
\ee
and $V_1$ accounts for the paramagnetic contribution to the current and $V_2$ to 
the diamagnetic one:
\be
V_1=-\frac{iet_\text{TB}}{\hbar} \sum_{n,i} \mathbf{A}_n\cdot\boldsymbol{\delta}_i \,\hat{a}^\dagger_{\mathbf{R}_n}\hat{b}_{\mathbf{R}_n+\boldsymbol{\delta}_i}+\text{h.c.},
\ee
\be
V_2=-\frac{e^2t_\text{TB}}{2\hbar^2} \sum_{n,i} \left(\mathbf{A}_n\cdot\boldsymbol{\delta}_i\right)^2 \hat{a}^\dagger_{\mathbf{R}_n}\hat{b}_{\mathbf{R}_n+\boldsymbol{\delta}_i}+\text{h.c.}\,.
\ee
In the following we will neglect the diamagnetic term $V_2$ as it only contributes to the Drude conductivity (we will return to this point later in the Appendix).

Using the momentum basis (\ref{mombasis}), we can write $V_1$ as:
\be
V_1= \frac{ et_\text{TB}}{iN_c\hbar} \sum_{\mathbf{k},\mathbf{k}^\prime,i} \left(\sum_n \mathbf{A}_n e^{ i(\mathbf{k}^\prime-\mathbf{k})\cdot\mathbf{R}_n}\right) \cdot\boldsymbol{\delta}_i e^{i\mathbf{k}^\prime\cdot\boldsymbol{\delta}_i}\hat{a}^\dagger_\mathbf{k}\hat{b}_\mathbf{k^\prime}+\text{h.c.} 
\ee
and defining the Fourier transform of the vector potential:
\be
\mathbf{A}(\mathbf{q})=\frac{1}{N_c} \sum_n e^{-i\mathbf{q}\cdot\mathbf{R}_n} \mathbf{A}_n,
\ee
and making the change of variables $\mathbf{q}=\mathbf{k}-\mathbf{k}^\prime$
it follows that:
\be
V_1=-\frac{iet_\text{TB}}{ \hbar} \sum_{\mathbf{k},\mathbf{q}}\sum_i \mathbf{A}(\mathbf{q})\cdot\boldsymbol{\delta}_i e^{i\left(\mathbf{k}-\frac{\mathbf{q}}{2}\right)\cdot\boldsymbol{\delta}_i}\hat{a}^\dagger_\mathbf{k+\mathbf{q}/2}\hat{b}_{\mathbf{k}-\mathbf{q}/2}+\text{h.c.}\,.
\ee
Using the identity $\mathbf{A}(\mathbf{q})=\mathbf{A}^*(-\mathbf{q})$ (since $\mathbf{A}(\mathbf{r})$ is real) we obtain:
\bea
V_1&=&-\frac{iet_\text{TB}}{ \hbar} \sum_{\mathbf{k},\mathbf{q}}\sum_i \mathbf{A}(\mathbf{q})\cdot\boldsymbol{\delta}_i\Big( e^{i\left(\mathbf{k}-\mathbf{q}/2\right)\cdot\boldsymbol{\delta}_i}\hat{a}^\dagger_\mathbf{k+\mathbf{q}/2}\hat{b}_{\mathbf{k}-\frac{\mathbf{q}}{2}}- \nonumber \\  
&-&e^{-i\left(\mathbf{k}+\mathbf{q}/2\right)\cdot\boldsymbol{\delta}_i}\hat{b}^\dagger_\mathbf{k+\mathbf{q}/2}\hat{a}_{\mathbf{k}-\mathbf{q}/2} \Big),
\eea
or
\bea
V_1&=&\frac{et_\text{TB}}{ \hbar} \sum_{\mathbf{k},\mathbf{q}} \mathbf{A}(\mathbf{q})\cdot \Big( \nabla_\mathbf{k}\phi_{\mathbf{k}-\mathbf{q}/2}\hat{a}^\dagger_\mathbf{k+\mathbf{q}/2}\hat{b}_{\mathbf{k}-\mathbf{q}/2}+ \nonumber \\  &+&\nabla_\mathbf{k} \phi^*_{\mathbf{k}+\mathbf{q}/2}\hat{b}^\dagger_\mathbf{k+\mathbf{q}/2}\hat{a}_{\mathbf{k}-\mathbf{q}/2} \Big),
\eea
now we use the basis that diagonalizes the Hamiltonian $H_0$ (\ref{bogtr}) with the choice of global phase $\varphi_\mathbf{k}=0$; thus it follows:
\bea
V_1&=&e\sum_{\mathbf{k},\mathbf{q}} \mathbf{A}(\mathbf{q})\cdot \boldsymbol{v}^\text{intra}_{\mathbf{k},\mathbf{q}}\left(\hat{c}^\dagger_{\mathbf{k}+\mathbf{q}/2}\hat{c}_{\mathbf{k}-\mathbf{q}/2}-\hat{d}^\dagger_{\mathbf{k}+\mathbf{q}/2}\hat{d}_{\mathbf{k}-\mathbf{q}/2}\right)+\nonumber\\
&+&\mathbf{A}(\mathbf{q})\cdot \boldsymbol{v}^\text{inter}_{\mathbf{k},\mathbf{q}}\left(\hat{c}^\dagger_{\mathbf{k}+\mathbf{q}/2}\hat{d}_{\mathbf{k}-\mathbf{q}/2}-\hat{d}^\dagger_{\mathbf{k}+\mathbf{q}/2}\hat{c}_{\mathbf{k}-\mathbf{q}/2}\right),
\eea
where we have defined:
\be
E_\mathbf{k}=t_\text{TB}|\phi_\mathbf{k}|,
\ee
\be
\boldsymbol{v}^\text{inter}_{\mathbf{k},\mathbf{q}}=\frac{a_0t_\text{TB}}{2\hbar} \left( e^{-i\Theta_{\mathbf{k}-\mathbf{q}/2}}\boldsymbol{\nabla}_\mathbf{k} \phi_{\mathbf{k}-\mathbf{q}/2} -e^{i\Theta_{\mathbf{k}+\mathbf{q}/2}}\boldsymbol{\nabla}_\mathbf{k} \phi^*_{\mathbf{k}+\mathbf{q}/2}\right),
\ee
\be
\boldsymbol{v}^\text{intra}_{\mathbf{k},\mathbf{q}}=\frac{a_0t_\text{TB}}{2\hbar} \left( e^{-i\Theta_{\mathbf{k}-\mathbf{q}/2}}\boldsymbol{\nabla}_\mathbf{k} \phi_{\mathbf{k}-\mathbf{q}/2} +e^{i\Theta_{\mathbf{k}+\mathbf{q}/2}}\boldsymbol{\nabla}_\mathbf{k} \phi^*_{\mathbf{k}+\mathbf{q}/2}\right)\,.
\ee

Using the Hamiltonian (\ref{eq_HV1V2}) in Eq. (\ref{heq}), neglecting the contribution
of the term $V_2$, and taking the expectation value of the resulting equation   for finite
$\mathbf{k}$ and $\mathbf{q}$
we arrive at the coupled integral equations (Bloch equations) for the elements of
the density matrix:
\begin{widetext}
\begin{subequations}
\bea
\left(i\partial_t-\omega^\text{intra}_{\mathbf{k},\mathbf{q}}-i\gamma_0 \right)n^c_{\mathbf{k},\mathbf{q}}=-i\gamma_0 f_\mathbf{k}^c \delta_{\mathbf{q},\boldsymbol{0}}+\sum_{\mathbf{q}^\prime} \left\{-\Omega^\text{intra}_{\mathbf{k}-\frac{1}{2}(\mathbf{q}+\mathbf{q}^\prime),\mathbf{q}^\prime}n^c_{\mathbf{k}-\frac{1}{2}\mathbf{q}^\prime,\mathbf{q}+\mathbf{q}^\prime}
+\Omega^\text{intra}_{\mathbf{k}+\frac{1}{2}(\mathbf{q}+\mathbf{q}^\prime),\mathbf{q}^\prime} n^c_{\mathbf{k}+\frac{1}{2}\mathbf{q}^\prime,\mathbf{q}+\mathbf{q}^\prime}-\right.\nonumber\\ \left.
-\Omega^\text{inter}_{\mathbf{k}-\frac{1}{2}(\mathbf{q}+\mathbf{q}^\prime),\mathbf{q}^\prime}p^{cv}_{\mathbf{k}-\frac{1}{2}\mathbf{q}^\prime,\mathbf{q}+\mathbf{q}^\prime}-\Omega^\text{inter}_{\mathbf{k}+\frac{1}{2}(\mathbf{q}+\mathbf{q}^\prime),\mathbf{q}^\prime} p^{vc}_{\mathbf{k}+\frac{1}{2}\mathbf{q}^\prime,\mathbf{q}+\mathbf{q}^\prime} \right\},
\eea
\bea
\left(i\partial_t+\omega^\text{intra}_{\mathbf{k},\mathbf{q}}-i\gamma_0 \right)n^v_{\mathbf{k},\mathbf{q}}=-i\gamma_0 f_\mathbf{k}^v \delta_{\mathbf{q},\boldsymbol{0}}+\sum_{\mathbf{q}^\prime} \left\{+\Omega^\text{intra}_{\mathbf{k}-\frac{1}{2}(\mathbf{q}+\mathbf{q}^\prime),\mathbf{q}^\prime}n^v_{\mathbf{k}-\frac{1}{2}\mathbf{q}^\prime,\mathbf{q}+\mathbf{q}^\prime}
-\Omega^\text{intra}_{\mathbf{k}+\frac{1}{2}(\mathbf{q}+\mathbf{q}^\prime),\mathbf{q}^\prime} n^v_{\mathbf{k}+\frac{1}{2}\mathbf{q}^\prime,\mathbf{q}+\mathbf{q}^\prime}
+\right.\nonumber\\ \left.+\Omega^\text{inter}_{\mathbf{k}-\frac{1}{2}(\mathbf{q}+\mathbf{q}^\prime),\mathbf{q}^\prime}p^{vc}_{\mathbf{k}-\frac{1}{2}\mathbf{q}^\prime,\mathbf{q}+\mathbf{q}^\prime} +\Omega^\text{inter}_{\mathbf{k}+\frac{1}{2}(\mathbf{q}+\mathbf{q}^\prime),\mathbf{q}^\prime} p^{cv}_{\mathbf{k}+\frac{1}{2}\mathbf{q}^\prime,\mathbf{q}+\mathbf{q}^\prime} \right\},
\eea
\bea
\left(i\partial_t-\omega^\text{inter}_{\mathbf{k},\mathbf{q}}-i\gamma_p \right)p^{cv}_{\mathbf{k},\mathbf{q}}=\sum_{\mathbf{q}^\prime} \left\{-\Omega^\text{intra}_{\mathbf{k}-\frac{1}{2}(\mathbf{q}+\mathbf{q}^\prime),\mathbf{q}^\prime}p^{cv}_{\mathbf{k}-\frac{1}{2}\mathbf{q}^\prime,\mathbf{q}+\mathbf{q}^\prime}
+\Omega^\text{intra}_{\mathbf{k}+\frac{1}{2}(\mathbf{q}+\mathbf{q}^\prime),\mathbf{q}^\prime} p^{cv}_{\mathbf{k}+\frac{1}{2}\mathbf{q}^\prime,\mathbf{q}+\mathbf{q}^\prime}
+\right.\nonumber\\ \left.+\Omega^\text{inter}_{\mathbf{k}-\frac{1}{2}(\mathbf{q}+\mathbf{q}^\prime),\mathbf{q}^\prime}n^{c}_{\mathbf{k}-\frac{1}{2}\mathbf{q}^\prime,\mathbf{q}+\mathbf{q}^\prime}-\Omega^\text{inter}_{\mathbf{k}+\frac{1}{2}(\mathbf{q}+\mathbf{q}^\prime),\mathbf{q}^\prime} n^{v}_{\mathbf{k}+\frac{1}{2}\mathbf{q}^\prime,\mathbf{q}+\mathbf{q}^\prime} \right\},
\eea
\bea
\left(i\partial_t+\omega^\text{inter}_{\mathbf{k},\mathbf{q}}-i\gamma_p \right)p^{vc}_{\mathbf{k},\mathbf{q}}=\sum_{\mathbf{q}^\prime} \left\{-\Omega^\text{intra}_{\mathbf{k}-\frac{1}{2}(\mathbf{q}+\mathbf{q}^\prime),\mathbf{q}^\prime}p^{vc}_{\mathbf{k}-\frac{1}{2}\mathbf{q}^\prime,\mathbf{q}+\mathbf{q}^\prime},
+\Omega^\text{intra}_{\mathbf{k}+\frac{1}{2}(\mathbf{q}+\mathbf{q}^\prime),\mathbf{q}^\prime} p^{vc}_{\mathbf{k}+\frac{1}{2}\mathbf{q}^\prime,\mathbf{q}+\mathbf{q}^\prime}-\right.\nonumber\\ \left.
-\Omega^\text{inter}_{\mathbf{k}-\frac{1}{2}(\mathbf{q}+\mathbf{q}^\prime),\mathbf{q}^\prime}n^{v}_{\mathbf{k}-\frac{1}{2}\mathbf{q}^\prime,\mathbf{q}+\mathbf{q}^\prime} +\Omega^\text{inter}_{\mathbf{k}+\frac{1}{2}(\mathbf{q}+\mathbf{q}^\prime),\mathbf{q}^\prime} n^{c}_{\mathbf{k}+\frac{1}{2}\mathbf{q}^\prime,\mathbf{q}+\mathbf{q}^\prime} \right\},
\eea
\label{ugly}
\end{subequations}
\end{widetext}
where two phenomenological relaxation times have been introduced and 
where:
\begin{subequations}
\be
n_{\mathbf{k},\mathbf{q}}^c=\langle c^\dagger_{\mathbf{k}+\mathbf{q}/2}c_{\mathbf{k}-\mathbf{q}/2}\rangle, 
\ee
\be
n_{\mathbf{k},\mathbf{q}}^v=\langle d^\dagger_{\mathbf{k}+\mathbf{q}/2}d_{\mathbf{k}-\mathbf{q}/2}\rangle,
\ee
\be
p_{\mathbf{k},\mathbf{q}}^{cv}=\langle c^\dagger_{\mathbf{k}+\mathbf{q}/2}d_{\mathbf{k}-\mathbf{q}/2}\rangle,
\ee
\be
p_{\mathbf{k},\mathbf{q}}^{vc}=\langle d^\dagger_{\mathbf{k}+\mathbf{q}/2}c_{\mathbf{k}-\mathbf{q}/2}\rangle,
\ee
\be
\Omega^\text{inter/intra}_{\mathbf{k},\mathbf{q}}=e \mathbf{A}(\mathbf{q}) \cdot \boldsymbol{v}^\text{inter/intra}_{\mathbf{k},\mathbf{q}}/\hbar,
\ee
\be
\omega^\text{inter}_{\mathbf{k},\mathbf{q}}=E_{\mathbf{k}-\mathbf{q}/2}+E_{\mathbf{k}+\mathbf{q}/2},
\ee
\be
\omega^\text{intra}_{\mathbf{k},\mathbf{q}}=E_{\mathbf{k}-\mathbf{q}/2}-E_{\mathbf{k}+\mathbf{q}/2}.
\ee
\end{subequations}

We now consider a field given by a pumping component with frequency $\omega_p$ and null wavenumber and a probing component with frequency $\omega$ and wavenumber $\mathbf{q}_0$:
\be
\mathbf{A}_n(\mathbf{R}_n,t)=
\frac{1}{2}\left[
\mathbf{A}_\text{p}e^{i\omega_p t}+\mathbf{A}_0 e^{i (\omega t-\mathbf{q}_0\cdot\mathbf{R}_n)}+\text{h.c.}
\right]\,,
\ee
with the Fourier transform:
\be
\mathbf{A}(\mathbf{q})=\mathbf{A}_\text{p} \cos(\omega_pt) \delta_{\mathbf{q},0}+\frac{\mathbf{A}_0}{2}\left(\delta_{\mathbf{q}_0,\mathbf{q}}e^{-i\omega t} +\delta_{-\mathbf{q}_0,\mathbf{q}}e^{i\omega t}\right),
\ee

Next we simplify the set of Eqs. (\ref{ugly}). First we consider the effects of the pumping field on the new electronic distribution $n^c_{\mathbf{k},\mathbf{0}}$, in a similar way to what has been done in Sec. \ref{neqdm}, that is we neglect the effect of
$\mathbf{A}_0$ since it is much smaller than $\mathbf{A}_\text{p}$. Using the result that the electronic distribution
converges to a steady-state we find:
\begin{subequations}
\be 
n^c_{\mathbf{k},\mathbf{0}}= f_\mathbf{k}^c +i\tau_0\left\langle\Omega^\text{inter}_{\mathbf{k},\mathbf{0}}\left(p^{cv}_{\mathbf{k},\mathbf{0}}+p^{vc}_{\mathbf{k},\mathbf{0}} \right)\right\rangle_t,
\ee 
\be 
n^v_{\mathbf{k},\mathbf{0}}= f_\mathbf{k}^v -i\tau_0\left\langle\Omega^\text{inter}_{\mathbf{k},\mathbf{0}}\left(p^{cv}_{\mathbf{k},\mathbf{0}}+p^{vc}_{\mathbf{k},\mathbf{0}} \right)\right\rangle_t,
\ee 
\be
\left(i\partial_t-\omega^\text{inter}_{\mathbf{k},\mathbf{0}}+i\gamma_p \right)p^{cv}_{\mathbf{k},\mathbf{0}}= \Omega^\text{inter}_{\mathbf{k},\mathbf{0}}\left(n^{c}_{\mathbf{k},\mathbf{0}}- n^{v}_{\mathbf{k},\mathbf{0}} \right),
 \label{eq_pcv_k}
\ee
\be
\left(i\partial_t+\omega^\text{inter}_{\mathbf{k},\mathbf{0}}+i\gamma_p \right)p^{vc}_{\mathbf{k},\mathbf{0}}=
-\Omega^\text{inter}_{\mathbf{k},\mathbf{0}}\left(n^{v}_{\mathbf{k},\mathbf{0}} 
 -n^{c}_{\mathbf{k},\mathbf{0}} \right)\,.
 \label{eq_pvc_k}
\ee
\end{subequations}
Making the ansatz that the off-diagonal part of the $p^{vc}$ and $p^{cv}$ tensor oscillates with the same frequency $\omega_p$ it follows 
from Eqs. (\ref{eq_pcv_k}) and (\ref{eq_pvc_k}) that:
\be
p^{vc}_{\mathbf{k},\mathbf{0}}=A_1(\omega_p)e^{i\omega_p t}+B_1(\omega_p)e^{-i\omega_p t},
\ee
\be
p^{cv}_{\mathbf{k},\mathbf{0}}=A_2(\omega_p)e^{i\omega_p t}+B_2(\omega_p)e^{-i\omega_p t},
\ee
and
\begin{subequations}
\be
A_1=\frac{1}{2\hbar} \frac{e\mathbf{A}_\text{p} \cdot \boldsymbol{v}^\text{inter}_{\mathbf{k},\mathbf{0}}}{-\omega_p-\omega^\text{inter}_{\mathbf{k},\mathbf{0}}+i\gamma_p} \left(n^c_{\mathbf{k},\mathbf{0}}-n^v_{\mathbf{k},\mathbf{0}}\right),
\ee
\be
B_1=\frac{1}{2\hbar} \frac{e\mathbf{A}^*_\text{p} \cdot \boldsymbol{v}^\text{inter}_{\mathbf{k},\mathbf{0}}}{\omega_p-\omega^\text{inter}_{\mathbf{k},\mathbf{0}}+i\gamma_p} \left(n^c_{\mathbf{k},\mathbf{0}}-n^v_{\mathbf{k},\mathbf{0}}\right),
\ee
\be
A_2=\frac{1}{2\hbar} \frac{e\mathbf{A}_\text{p} \cdot \boldsymbol{v}^\text{inter}_{\mathbf{k},\mathbf{0}} }{-\omega_p+\omega^\text{inter}_{\mathbf{k},\mathbf{0}}+i\gamma_p} \left(n^c_{\mathbf{k},\mathbf{0}}-n^v_{\mathbf{k},\mathbf{0}}\right),
\ee
\be
B_2=\frac{1}{2\hbar} \frac{e\mathbf{A}^*_\text{p} \cdot \boldsymbol{v}^\text{inter}_{\mathbf{k},\mathbf{0}} }{\omega_p+\omega^\text{inter}_{\mathbf{k},\mathbf{0}}+i\gamma_p} \left(n^c_{\mathbf{k},\mathbf{0}}-n^v_{\mathbf{k},\mathbf{0}}\right),
\ee
\end{subequations}
and if we define:
\be
\beta_\mathbf{k}= \frac{ \tau_0 \tau_p\left|\Omega^\text{inter}(\omega)\right|^2\left(1+\tau_p^2\left(\omega^2+{\omega^\text{inter}_{\mathbf{k},\mathbf{0}}}^2 \right)\right)}{\tau_p^4(\omega^2-{\omega^\text{inter}_{\mathbf{k},\mathbf{0}}}^2)^2+2\tau_p^2(\omega^2+{\omega^\text{inter}_{\mathbf{k},\mathbf{0}} }^2)+1},
\ee
where:
\be
\tilde{\Omega}^\text{inter}_\mathbf{k}(\omega)=e\mathbf{A}_\text{p} \cdot \boldsymbol{v}^\text{inter}_{\mathbf{k},\mathbf{0}}/\hbar=\frac{ea_0 \omega^\text{inter}_{\mathbf{k},\mathbf{0}} \mathbf{E}\cdot\boldsymbol{\nabla}_\mathbf{k}\Theta_\mathbf{k}}{2\omega\hbar}
\ee
where $\tau_p=1/\gamma_p$, we have:
\be
n^c_\mathbf{k}-f^c_\mathbf{k}=\beta_\mathbf{k}(n^v_\mathbf{k}-n^c_\mathbf{k}),
\ee
\be
n^v_\mathbf{k}-f^v_\mathbf{k}=\beta_\mathbf{k}(n^c_\mathbf{k}-n^v_\mathbf{k}),
\ee
that are the equivalent of  Eqs. (\ref{steady_state_sol}) in the Couloumb gauge, except that the Rabi frequency, $\tilde{\Omega}^\text{inter}_\mathbf{k}(\omega)$, has a factor $\omega_\mathbf{k}/\omega$ with relation to Eq. (\ref{rabi_indep}): $\tilde{\Omega}^\text{inter}(\omega)=\omega_\mathbf{k}/\omega\,\bar{\Omega}_\mathbf{k}$.

Returning to the set of Eqs. (\ref{ugly}) we now consider the effect of the two 
fields. Since the pumping field has zero momentum it only contributes to the 
out-of-equilibrium distribution function calculated above. In practical terms this amounts
to replace the equilibrium distribution functions by the out-of-equilibrium ones 
for the optical conductivity due to the probe.
Following this procedures with the new electronic distribution we can calculate the electronic current. For $\mathbf{q}=\mathbf{q}_0$, we have 
\begin{subequations}
\be
\left(i\partial_t-\omega^\text{intra}_{\mathbf{k},\mathbf{q}_0}+i\gamma_0 \right)n^c_{\mathbf{k},\mathbf{q}_0}= \Omega^\text{intra}_{\mathbf{k},-\mathbf{q}_0}\left(n^c_{\mathbf{k}-\frac{\mathbf{q}_0}{2},\mathbf{0}}-n^c_{\mathbf{k}+\frac{\mathbf{q}_0}{2},\mathbf{0}}\right),
\ee
\be
\left(i\partial_t+\omega^\text{intra}_{\mathbf{k},\mathbf{q}_0}+i\gamma_0 \right)n^v_{\mathbf{k},\mathbf{q}_0}= \Omega^\text{intra}_{\mathbf{k},-\mathbf{q}_0}\left(n^v_{\mathbf{k}+\frac{\mathbf{q}_0}{2},\mathbf{0}}
- n^v_{\mathbf{k}-\frac{\mathbf{q}_0}{2},\mathbf{0}}\right),
\ee
\be
\left(i\partial_t-\omega^\text{inter}_{\mathbf{k},\mathbf{q}_0}+i\gamma_p \right)p^{cv}_{\mathbf{k},\mathbf{q}_0}= 
 \Omega^\text{inter}_{\mathbf{k},-\mathbf{q}_0}\left(n^{c}_{\mathbf{k}+\frac{\mathbf{q}_0}{2},\mathbf{0}} - n^{v}_{\mathbf{k}-\frac{\mathbf{q}_0}{2},\mathbf{0}}\right),
\ee
\be
\left(i\partial_t+\omega^\text{inter}_{\mathbf{k},\mathbf{q}_0}+i\gamma_p \right)p^{vc}_{\mathbf{k},\mathbf{q}_0}=
\Omega^\text{inter}_{\mathbf{k},-\mathbf{q}_0}\left( n^{c}_{\mathbf{k}-\frac{\mathbf{q}_0}{2},\mathbf{0}}-n^{v}_{\mathbf{k}+\frac{\mathbf{q}_0}{2},\mathbf{0}}\right)\,.
\ee \label{apeq3x}
\end{subequations}
In the rotating wave approximation Eq. (\ref{apeq3x}) becomes:
\begin{subequations}
\be 
n^c_{\mathbf{k},-\mathbf{q}_0}=-e\frac{\mathbf{A}_0\cdot  \boldsymbol{v}^\text{intra}_{\mathbf{k}, \mathbf{q}_0}}{2\hbar} \frac{n^c_{\mathbf{k}-\frac{\mathbf{q}_0}{2},\mathbf{0}}
- n^c_{\mathbf{k}+\frac{\mathbf{q}_0}{2},\mathbf{0}}}{ \omega-\omega^\text{intra}_{\mathbf{k},-\mathbf{q}_0}+i\gamma_0  }e^{-i\omega t},
\ee 
\be 
n^v_{\mathbf{k},-\mathbf{q}_0}= e\frac{\mathbf{A}_0\cdot  \boldsymbol{v}^\text{intra}_{\mathbf{k}, \mathbf{q}_0}}{2\hbar}\frac{n^v_{\mathbf{k}-\frac{\mathbf{q}_0}{2},\mathbf{0}}
- n^v_{\mathbf{k}+\frac{\mathbf{q}_0}{2},\mathbf{0}}}{ \omega+\omega^\text{intra}_{\mathbf{k},-\mathbf{q}_0}+i\gamma_0  }e^{ i\omega t},
\ee 
\be 
p^{cv}_{\mathbf{k},-\mathbf{q}_0}= 
 e\frac{\mathbf{A}_0\cdot  \boldsymbol{v}^\text{inter}_{\mathbf{k}, \mathbf{q}_0}}{2\hbar}\frac{n^{c}_{\mathbf{k}-\frac{\mathbf{q}_0}{2},\mathbf{0}} - n^{v}_{\mathbf{k}+\frac{\mathbf{q}_0}{2},\mathbf{0}}}{ \omega-\omega^\text{inter}_{\mathbf{k},-\mathbf{q}_0}+i\gamma_p  }e^{-i\omega t},
\ee 
\be 
p^{vc}_{\mathbf{k},-\mathbf{q}_0}=
-e\frac{\mathbf{A}_0\cdot  \boldsymbol{v}^\text{inter}_{\mathbf{k}, \mathbf{q}_0}}{2\hbar}\frac{ n^{v}_{\mathbf{k}-\frac{\mathbf{q}_0}{2},\mathbf{0}}- n^{c}_{\mathbf{k}+\frac{\mathbf{q}_0}{2},\mathbf{0}}}{ \omega+\omega^\text{inter}_{\mathbf{k},-\mathbf{q}_0}+i\gamma_p  }e^{ i\omega t} ,
\ee  \label{apeq4x}
\end{subequations}

We can write the current in the $n$th cell:
\bea
\hat{\mathbf{j}}_n = -\frac{\delta V_1}{\delta \mathbf{A}_n}&=&\frac{e t_\text{TB}}{N_c} \sum_{\mathbf{k},\mathbf{k}^\prime,i} \boldsymbol{\delta}_i \Big[e^{-i(\mathbf{k}-\mathbf{k}^\prime)\cdot\mathbf{R}_n}e^{i\mathbf{k}^\prime\cdot\boldsymbol{\delta}_i}\hat{a}_\mathbf{k}^\dagger \hat{b}_{\mathbf{k}^\prime}+ \nonumber \\  &+& e^{i(\mathbf{k}-\mathbf{k}^\prime)\cdot \mathbf{R}_n}e^{-i\mathbf{k}^\prime \cdot \boldsymbol{\delta}_i} \hat{b}^\dagger_{\mathbf{k}^\prime} \hat{a}_\mathbf{k}\Big],
\label{eq_j_op}
\eea
after making the Bogoliubov transformation (\ref{bogtr}) 
in Eq. (\ref{eq_j_op})
and taking the expectation value we have:
\be 
\mathbf{J}_n=\frac{e}{\hbar N_c} \sum_{\mathbf{k},\mathbf{q}} e^{i\mathbf{q}\cdot\mathbf{R}_n} \boldsymbol{v}^\text{intra}_{\mathbf{k},\mathbf{q}}\left(n^c_{\mathbf{k},\mathbf{q}}-n^v_{\mathbf{k},\mathbf{q}}\right)
- \boldsymbol{v}^\text{inter}_{\mathbf{k},\mathbf{q}}\left(p^{cv}_{\mathbf{k},\mathbf{q}}-p^{vc}_{\mathbf{k},\mathbf{q}}\right)\,.
\ee 
The current up to first order in the probing field reads:
\be
\mathbf{J}_n(\mathbf{R}_n,t)=\frac{1}{2}\left(\mathbf{J}_0 e^{i (\omega t-\mathbf{q}_0\cdot\mathbf{R}_n)}+\text{h.c.}\right),
\ee
and finally the current $\mathbf{J}_0$ can be written as:
\begin{widetext}
\bea
\mathbf{J}_0 &=& \frac{2e^2 }{\hbar N_c}\sum_\mathbf{k}\Bigg\{ - \boldsymbol{v}^\text{intra}_{\mathbf{k},-\mathbf{q}_0}\frac{\mathbf{A}_0\cdot \boldsymbol{v}^\text{intra}_{\mathbf{k}, \mathbf{q}_0}\left(n^c_{\mathbf{k}-\frac{1}{2}\mathbf{q}_0,\mathbf{0}}
- n^c_{\mathbf{k}+\frac{1}{2}\mathbf{q}_0,\mathbf{0}}\right)}{\omega-\omega^\text{intra}_{\mathbf{k},-\mathbf{q}_0}+i\gamma_0 } -  \boldsymbol{v}^\text{intra}_{\mathbf{k},-\mathbf{q}_0}\frac{\mathbf{A}_0\cdot  \boldsymbol{v}^\text{intra}_{\mathbf{k}, \mathbf{q}_0}\left(n^v_{\mathbf{k}-\frac{1}{2}\mathbf{q}_0,\mathbf{0}}
- n^v_{\mathbf{k}+\frac{1}{2}\mathbf{q}_0,\mathbf{0}}\right)}{\omega+\omega^\text{intra}_{\mathbf{k},-\mathbf{q}_0}+i\gamma_0 }
+ \nonumber \\  
 &+&  \boldsymbol{v}^\text{inter}_{\mathbf{k},-\mathbf{q}_0}
\frac{ \mathbf{A}_0\cdot  \boldsymbol{v}^\text{inter}_{\mathbf{k}, \mathbf{q}_0} \left(n^{c}_{\mathbf{k}-\frac{1}{2}\mathbf{q}_0,\mathbf{0}} - n^{v}_{\mathbf{k}+\frac{1}{2}\mathbf{q}_0,\mathbf{0}}\right)}{\omega-\omega^\text{inter}_{\mathbf{k},-\mathbf{q}_0}+i\gamma_p }+     \boldsymbol{v}^\text{inter}_{\mathbf{k},-\mathbf{q}_0}
\frac{\mathbf{A}_0\cdot  \boldsymbol{v}^\text{inter}_{\mathbf{k}, \mathbf{q}_0}\left( n^{v}_{\mathbf{k}-\frac{1}{2}\mathbf{q}_0,\mathbf{0}}- n^{c}_{\mathbf{k}+\frac{1}{2}\mathbf{q}_0,\mathbf{0}}\right)}{\omega+\omega^\text{inter}_{\mathbf{k},-\mathbf{q}_0}+i\gamma_p}
\Bigg\},
\eea
\end{widetext}
where we included a factor of 2 to account for the spin degeneracy. The relation between the electric field and the potential vector in frequency space is: 
\be
\mathbf{A}_0= \frac{\mathbf{E}_0}{i\omega}\,.
\ee
We can finally write the conductivity tensor as:
\be
\sigma_{ij}^\text{inter}(\mathbf{q},\omega) = \frac{2ie^2}{\hbar\omega S} \sum_{\mathbf{k}, \lambda=\pm} \lambda\frac{n^v_{\mathbf{k}+\lambda\mathbf{q}/2}-n^c_\mathbf{k-\lambda\mathbf{q}/2}}{\omega-\lambda \omega^\text{inter}_{\mathbf{k},\mathbf{q}}+i\gamma_p} {v_i}^\text{inter}_{\mathbf{k},\mathbf{q}} {v_j}^\text{inter}_{\mathbf{k},-\mathbf{q}},
\label{eq_sigma_inter}
\ee
\be
\sigma_{ij}^\text{intra}(\mathbf{q},\omega) = \frac{2ie^2}{\hbar\omega S} \sum_{\mathbf{k}, \lambda=\pm} \frac{n^\lambda_{\mathbf{k}-\mathbf{q}/2}
-n^\lambda_\mathbf{k+\mathbf{q}/2}}{\omega-\lambda \omega^\text{intra}_{\mathbf{k},\mathbf{q}}+i\gamma_0} {v_i}^\text{intra}_{\mathbf{k},\mathbf{q}} {v_j}^\text{intra}_{\mathbf{k},-\mathbf{q}},
\label{eq_sigma_intra}
\ee
where we defined $S=N_c a_0^2$ with $N_c$ the number of unit cells in the crystal and $n^\lambda_{\mathbf{k}}=n^\lambda_{\mathbf{k},\mathbf{0}} $ is expected value of the diagonal element of the density matrix, with $\lambda=+$($-$) the conductance (valence) band. Introducing the relation
\begin{equation}
\frac{1}{\bar\omega(\bar\omega-\lambda \omega_{\mathbf{k},\mathbf{q}})}=-
\frac{1}{\bar\omega\lambda \omega_{\mathbf{k},\mathbf{q}}}+
\frac{1}{\lambda \omega_{\mathbf{k},\mathbf{q}}(\bar\omega-\lambda \omega_{\mathbf{k},\mathbf{q}})}
\end{equation}
in Eqs. (\ref{eq_sigma_inter}) and (\ref{eq_sigma_intra}),
where $\bar\omega=\omega+i\gamma$, 
we find that the term proportional to $-1/\bar\omega$ has exactly the same form 
---that is, proportional to $1/\bar\omega$---  as the  diamagnetic term we have ignored, and therefore these two terms should grouped. This procedure allows the determination the regular part of the conductivity.
Finally, we have for the regular part the conductivity tensor the results:
\be
\sigma_{ij}^\text{R,inter}(\mathbf{q},\omega) = \frac{2ie^2}{\hbar S} \sum_{\mathbf{k}, \lambda=\pm} \frac{(n^c_\mathbf{k+\lambda\mathbf{q}/2}-n^v_{\mathbf{k}-\lambda\mathbf{q}/2}) {v_i}^\text{inter}_{\mathbf{k},\mathbf{q}} {v_j}^\text{inter}_{\mathbf{k},-\mathbf{q}} }{ \omega^\text{inter}_{\mathbf{k},\mathbf{q}}(\omega-\lambda \omega^\text{inter}_{\mathbf{k},\mathbf{q}}+i\gamma_p)} ,
\label{eq_sigma_inter_f}
\ee
\be
\sigma_{ij}^\text{R,intra}(\mathbf{q},\omega) = \frac{2ie^2}{\hbar S} \sum_{\mathbf{k}, \lambda=\pm} \frac{  (n^\lambda_\mathbf{k-\mathbf{q}/2}-n^\lambda	_{\mathbf{k}+\mathbf{q}/2}){v_i}^\text{intra}_{\mathbf{k},\mathbf{q}} {v_j}^\text{intra}_{\mathbf{k},-\mathbf{q}} }{\lambda \omega^\text{intra}_{\mathbf{k},\mathbf{q}}(\omega-\lambda \omega^\text{intra}_{\mathbf{k},\mathbf{q}}+i\gamma_0)}\,,
\label{eq_sigma_intra _f}
\ee
whose both real and imaginary parts are not divergent when $\omega\rightarrow0$. The divergent piece of the conductivity when $\omega\rightarrow0$ is associated to the imaginary part of the neglected contributions and is nothing but the Drude conductivity. Note that 
when $\mathbf{q}\rightarrow0$, $\sigma_{ij}^\text{intra}(\mathbf{q},\omega)$ is minus  the neglected term and therefore they cancel
each other in that limit.  

Let us now return to the contribution to the total conductivity
of the term  $V_2$, termed diamagnetic contribution. Following the same steps that we used to calculate the
paramagnetic term due to $V_1$, we find that the diamagnetic current in the 
$n$th unit cell is given by:
\be
\hat{\mathbf{j}}_n^\text{dia}=-\frac{\delta V_2}{\delta\mathbf{A}_n}=\frac{e^2 t_\text{TB}}{\hbar^2} \sum_{i} \left(\mathbf{A}_n\cdot\boldsymbol{\delta}_i \right)\boldsymbol{\delta}_i \hat{a}^\dagger_{\mathbf{R}_n} \hat{b}_{\mathbf{R}_n+\boldsymbol{\delta}_i}+\text{h.c.}\,.
\ee
After using the Fourier representation of the creation and annihilation operators, and the Bogoliubov transformation that diagonalizes $H_0$, we find:
\begin{widetext}
\bea
\hat{\mathbf{j}}_n^\text{dia}&=&\frac{e^2 t_\text{TB}}{2\hbar^2 N_c} \sum_{\mathbf{k},\mathbf{q}}\left(\sum_i \boldsymbol{\delta}_i (\mathbf{A}_n \cdot \boldsymbol{\delta}_i) e^{i(\mathbf{k}+\mathbf{q}/2)\cdot\boldsymbol{\delta}_i} \right)e^{-i\Theta_{\mathbf{k}-\mathbf{q}/2}} e^{i\mathbf{q} \cdot \mathbf{R}_n}\nonumber \\ &\Bigg\{& \hat{c}^\dagger_{\mathbf{k}+\mathbf{q}/2}\hat{c}^\dagger_{\mathbf{k}-\mathbf{q}/2} 
-\hat{d}^\dagger_{\mathbf{k}+\mathbf{q}/2}\hat{d}^\dagger_{\mathbf{k}-\mathbf{q}/2} 
-\hat{c}^\dagger_{\mathbf{k}+\mathbf{q}/2}\hat{d}^\dagger_{\mathbf{k}-\mathbf{q}/2} 
+\hat{d}^\dagger_{\mathbf{k}+\mathbf{q}/2}\hat{c}^\dagger_{\mathbf{k}-\mathbf{q}/2} \Bigg\}+\text{h.c.}\,.
\eea
\end{widetext}
Next we take the expectation value of the previous equation, insert a factor $2$ to account for the spin degeneracy, truncate at first order in the field $\mathbf{A}_n$ the expectation value,  and perform a summation in $\mathbf{q}$, obtaining:
\bea
\mathbf{J}^\text{dia}_n &=& \frac{ 2e^2 t_\text{TB}}{\hbar^2 N_c} \sum_\mathbf{k}\left( \sum_i \frac{1}{2}e^{i\mathbf{k}\cdot\boldsymbol{\delta}_i}\boldsymbol{\delta}_i(\mathbf{A}_n\cdot\boldsymbol{\delta}_i) e^{-i\Theta_\mathbf{k}}+\text{h.c.} \right)\nonumber\\ 
&& \times\left(n^c_{\mathbf{k},\mathbf{0}}-n^v_{\mathbf{k},\mathbf{0} }\right)\,.
\eea
The term inside the first pair of braces can be rewritten with the aid of the following 
relation:
\be
X_{ij}=\frac{1}{2}t_\text{TB}\sum_l e^{i\mathbf{k}\cdot\boldsymbol{\delta}_l} \boldsymbol{\delta}_l\cdot \mathbf{u}_i \left(\boldsymbol{\delta}_l \cdot \mathbf{u}_j \right) e^{-i\Theta_\mathbf{k}}+\text{h.c.}, \label{X_def}
\ee
with $\mathbf{u}_i$ a versor in the direction $i=x,y$. The function (\ref{X_def}) can be rewritten as [note that $\mathbf{k}$ in Eq. (\ref{X_def}) is dimensionful and therefore the derivatives in order to the
components of $\mathbf{k}$ are also dimensionful] :
\be
X_{ij}(\mathbf{k})=  t_\text{TB}a_0^2\Re\left[e^{-i\Theta_\mathbf{k}}\partial_i \partial_j \phi(\mathbf{k})\right].
\ee
Finally the diamagnetic term of the conductivity read as:
\be
\sigma_{ij}^\text{dia}(\omega)=-\frac{2ie^2}{\hbar^2 S (\omega+i\gamma_p)} \sum_\mathbf{k} X_{ij}(\mathbf{k}) \left( n^c_{\mathbf{k},0}-n^v_{\mathbf{k},0}\right).
\ee
Putting all together, we have  the Drude term for finite $\mathbf{q}$:
\begin{equation}
\sigma_{ij}^\text{D}(\mathbf{q},\omega)= \sigma_{ij}^\text{dia}(\omega)
+\sigma_{ij}^\text{D,intra}(\mathbf{q},\omega)+
\sigma_{ij}^\text{D,inter}(\mathbf{q},\omega)\,,
\end{equation} 
where
\begin{align}
\sigma_{ij}^\text{D,intra}(\mathbf{q},\omega)&=
-\frac{2ie^2}{\hbar S} \sum_{\mathbf{k}, \lambda=\pm} \frac{  (n^\lambda_\mathbf{k-\mathbf{q}/2}-n^\lambda_{\mathbf{k}+\mathbf{q}/2}){v_i}^\text{intra}_{\mathbf{k},\mathbf{q}} {v_j}^\text{intra}_{\mathbf{k},-\mathbf{q}} }{\lambda \omega^\text{intra}_{\mathbf{k},\mathbf{q}}(\omega+i\gamma_0)}\,,
\\
\sigma_{ij}^\text{D,inter}(\mathbf{q},\omega)&=-
\frac{2ie^2}{\hbar S} \sum_{\mathbf{k}, \lambda=\pm} \frac{(n^v_\mathbf{k+\lambda\mathbf{q}/2}-n^c_{\mathbf{k}-\lambda\mathbf{q}/2}) {v_i}^\text{inter}_{\mathbf{k},\mathbf{q}} {v_j}^\text{inter}_{\mathbf{k},-\mathbf{q}} }{ \omega^\text{inter}_{\mathbf{k},\mathbf{q}}(\omega+i\gamma_p)}\,. 
\end{align}
The total conductivity is the sum of all the three terms:
$\sigma_{ij}(\mathbf{q},\omega)=\sigma_{ij}^\text{D}(\mathbf{q},\omega)+\sigma_{ij}^\text{R,inter}(\mathbf{q},\omega)+\sigma_{ij}^\text{R,intra}(\mathbf{q},\omega)$.
Let us stress again that in the limit $\mathbf{q}\rightarrow0$
we have $\sigma_{ij}^\text{D,intra}(\mathbf{q}\rightarrow0,\omega)=-\sigma_{ij}^\text{R,intra}(\mathbf{q}\rightarrow0,\omega)$, in which case the Drude conductivity is given solely by
\begin{equation}
\sigma_{ij}^\text{D}(0,\omega)=\sigma_{ij}^\text{dia}(\omega)+
\sigma_{ij}^\text{D,inter}(0,\omega)\,.
\end{equation}
This result concludes the discussion of the optical response of graphene
within  the tight-binding approximation.
\end{appendix}

\bibliographystyle{apsrev4-1}
\bibliography{references}
 
\end{document}